\documentclass[journal]{IEEEtran}
\usepackage[T1]{fontenc}
\usepackage{cite}
\usepackage{amsmath,amssymb,amsfonts}
\usepackage{algorithm}
\usepackage{algorithmic}
\usepackage{graphicx}
\usepackage{textcomp}
\usepackage{xcolor}
\usepackage[normalem]{ulem} 
\usepackage{subfigure}
\usepackage{booktabs}
\usepackage{multirow}
\usepackage{diagbox}
\usepackage{url}
\usepackage{threeparttable}
\usepackage{colortbl}
\usepackage{array}
\usepackage{textcase}
\usepackage{rotating}
\usepackage{etoolbox}

\makeatletter
\def\UrlAlphabet{%
      \do\a\do\b\do\c\do\d\do\e\do\f\do\g\do\h\do\i\do\j%
      \do\k\do\l\do\m\do\n\do\o\do\p\do\q\do\r\do\s\do\t%
      \do\u\do\v\do\w\do\x\do\y\do\z\do\A\do\B\do\C\do\D%
      \do\E\do\F\do\G\do\H\do\I\do\J\do\K\do\L\do\M\do\N%
      \do\O\do\P\do\Q\do\R\do\S\do\T\do\U\do\V\do\W\do\X%
      \do\Y\do\Z}
\def\UrlDigits{\do\1\do\2\do\3\do\4\do\5\do\6\do\7\do\8\do\9\do\0}
\g@addto@macro{\UrlBreaks}{\UrlOrds}
\g@addto@macro{\UrlBreaks}{\UrlAlphabet}
\g@addto@macro{\UrlBreaks}{\UrlDigits}

\patchcmd{\@algocf@start}
  {-1.5em}
  {0pt}
  {}{}

\makeatother

\newcommand{\Times}[2]{${\text{#1}\times\text{#2}}$}
\newcommand{\SNR}[2]{${\text{SNR} #1 \text{#2}~\text{dB}}$}

\graphicspath{{figures/}}

\def\nt{N_{\rm t}}
\def\nr{N_{\rm r}}

\begin{document}

\setlength{\textfloatsep}{5pt}  
\setlength{\floatsep}{5pt}
\ifdefined \GramaCheck
  \newcommand{\CheckRmv}[1]{}
  \newcommand{\figref}[1]{Figure 1}%
  \newcommand{\tabref}[1]{Table 1}%
  \newcommand{\secref}[1]{Section 1}
  \newcommand{\algref}[1]{Algorithm 1}
  \renewcommand{\eqref}[1]{Equation 1}
\else
  \newcommand{\CheckRmv}[1]{#1}
  \newcommand{\figref}[1]{Fig.~\ref{#1}}%
  \newcommand{\tabref}[1]{Table~\ref{#1}}%
  \newcommand{\secref}[1]{Sec.~\ref{#1}}
  \newcommand{\algref}[1]{Algorithm~\ref{#1}}
  \renewcommand{\eqref}[1]{(\ref{#1})}
\fi
\newtheorem{theorem}{Theorem}
\newtheorem{proposition}{Proposition}
\newtheorem{assumption}{Assumption}
\newtheorem{definition}{Definition}
\newtheorem{condition}{Condition}
\newtheorem{property}{Property}
\newtheorem{remark}{Remark}
\newtheorem{lemma}{Lemma}
\newtheorem{corollary}{Corollary}
%
\title{{A Graph Foundation Model for \\ Large-Scale MIMO Detection}}

%
%
\author{Xingyu~Zhou,~
        Le~Liang,~\IEEEmembership{Member,~IEEE,}
        Hao~Ye,~\IEEEmembership{Member,~IEEE,}
        Jing~Zhang,~\IEEEmembership{Member,~IEEE,}
        Chao-Kai~Wen,~\IEEEmembership{Fellow,~IEEE,}
        Xiao~Li,~\IEEEmembership{Member,~IEEE,}
        Shi~Jin,~\IEEEmembership{Fellow,~IEEE,}
        and~Wei~Zhang,~\IEEEmembership{Fellow,~IEEE}
\thanks{X.~Zhou, L. Liang, J.~Zhang, X. Li, and S.~Jin are with the National
Mobile Communications Research Laboratory, Southeast University, Nanjing 210096, China (e-mail: \protect \url{xy_zhou@seu.edu.cn}; lliang@seu.edu.cn; jingzhang@seu.edu.cn; \protect \url{li_xiao@seu.edu.cn}; jinshi@seu.edu.cn).}
\thanks{H. Ye is with the Department of Electrical and Computer Engineering, University of California, Santa Cruz, CA 95064, USA
(e-mail: hye30@ucsc.edu).}
\thanks{C.-K. Wen is with the Institute of Communications Engineering, National Sun Yat-sen University, Kaohsiung 80424, Taiwan 
(e-mail: chaokai.wen@mail.nsysu.edu.tw).}
\thanks{W. Zhang is with the School of Electrical Engineering and Telecommunications, University of New South Wales, Sydney, NSW 2052, Australia
(e-mail: w.zhang@unsw.edu.au).}
}

\maketitle

\begin{abstract}
Large-scale multiple-input multiple-output (MIMO) detection is fundamental to modern wireless networks but constrained by performance-complexity trade-offs.  Existing detectors, whether classical or learning-based, often fall short in either scalability or generalizability across heterogeneous scenarios. To overcome these limitations, we introduce a wireless-native graph foundation model (GFM) tailored for large-scale MIMO detection. The proposed GFM employs a physics-informed hybrid architecture, integrating the local correlation extraction of message passing neural networks with the global attention of graph Transformers, encoding the physical interference patterns from the expectation propagation algorithm. Via extensive pre-training, this synergy enables the learning of a general-purpose detection mapping scalable across antenna dimensions and channel conditions. For rapid downstream deployment, parameter-efficient fine-tuning is leveraged to adapt the GFM to specific non-ideal system regimes with minimal overhead. To enhance inference efficiency, a mixture-of-experts mechanism is embedded at downstream deployment to dynamically activate only the necessary sub-modules. Evaluations show that the proposed GFM consistently outperforms classical detectors and advanced data-driven baselines in accuracy, configuration generality, and cross-scenario transferability across various challenging zero-shot and few-shot conditions.
\end{abstract}

%
\IEEEpeerreviewmaketitle

\section{Introduction}  

\IEEEPARstart{M}{ultiple}-input multiple-output (MIMO) technology is fundamental to modern wireless communications, facilitating terabit-per-second (Tbps) data rates and ubiquitous connectivity \cite{wang2024tutorial}. These requirements of the sixth-generation (6G) wireless networks and beyond have driven the development of extremely large-scale MIMO (XL-MIMO), which scales the antenna count from 16--128 to 512 or more \cite{shafi2025industrial}. However, the antenna increase further intensifies the complexity of signal processing tasks that are already challenging under significant computational constraints, such as high-dimensional channel state information (CSI) acquisition \cite{zhou2025generative} and large-scale MIMO detection \cite{yangFiftyYearsMIMO2015}.

In particular, MIMO detection is a fundamental problem in MIMO communications, aiming to recover the transmitted symbol vector from the received signal and CSI \cite{yangFiftyYearsMIMO2015}. Optimal MIMO detection is known to be non-deterministic polynomial-time hard (NP-hard) due to the combinatorial nature of the search space. Although sphere decoders can reduce the average complexity compared to the exhaustive search of the optimal detector while maintaining near-optimal performance, they are still computationally prohibitive in large-scale MIMO systems \cite{agrellClosestPointSearch2002}. 
Therefore, a plethora of suboptimal algorithms have been proposed to strike a balance between complexity and performance, including linear minimum mean squared error (LMMSE), successive interference cancellation, and message-passing-based detectors \cite{yangFiftyYearsMIMO2015}. 
However, these algorithms often rely on idealized assumptions that may not hold in practical scenarios, leading to severe performance degradation. 

{Recent standardization studies \cite{3gpp38843study2023} mark a clear trend toward embedding artificial intelligence (AI) into the physical layer of wireless networks \cite{lin2026toward}. Instead of applying AI as an afterthought, this trend anticipates an \textit{AI-native} architecture \cite{hoydis2021toward} that shifts the physical-layer design from rigid mathematical algorithms to data-driven neural networks (NNs) capable of scenario-aware adaptation. For high-dimensional processing tasks like large-scale MIMO detection, {such adaptive intelligence can help address conventional challenges, including computational bottlenecks, non-ideal hardware conditions, and cross-scenario generalization.}}

Consequently, learning-based approaches have been increasingly explored to tackle the critical performance-complexity trade-off in MIMO detection. For example, deep unfolding architectures integrate learnable parameters into classical iterative detection processes to expedite convergence with limited training overhead \cite{samuelLearningDetect2019,khaniAdaptiveNeuralSignal2020,shlezinger2023model}. {In parallel,} neural augmentation approaches utilize architectures such as graph NNs (GNNs) to enhance the intermediate
computations of traditional detectors, relying heavily on data to learn complex interference patterns without rigid statistical assumptions \cite{kosasihGraphNeuralNetwork2022a,pratik2020re}.  
However, these existing approaches often struggle to generalize across different system configurations and out-of-distribution (OOD) channel environments, violating the intrinsic adaptive and autonomous ambitions of AI-native communications and limiting their practical applicability in dynamic wireless environments.

In recent years, research has begun to explore the potential of large-scale AI models, also known as \textit{foundation models} \cite{bommasani2021opportunities}, in wireless communications inspired by their unprecedented generalization capabilities in vision and language processing. 
Early investigations resort to adapting general-purpose foundation models, such as large language models (LLMs) and large vision models (LVMs), to physical-layer tasks \cite{liang2026large}, including channel prediction \cite{liu2024llm4cp}, beam management \cite{sheng2025beam}, and multi-task joint processing \cite{zheng2025large}. Despite their promising performance, there remains a significant modality gap between language/vision tokens and wireless data, along with the strict latency requirements of physical-layer processing that these pre-trained foundation models struggle to satisfy.
To address these mismatches, there is a growing interest in developing wireless-native foundation models that are pre-trained from scratch on massive wireless data to learn compact and generalizable representations for physical-layer tasks \cite{alikhani2024large,liu2025wifo,yang2025wirelessgpt,sheng2025wireless}, potentially leading to efficient and adaptable intelligence that is anticipated by AI-native communications. 
Nonetheless, the application of foundation models in the physical layer is still in its infancy, particularly for MIMO detection, where the construction of a foundation model that captures complex interference patterns and {generalizes across diverse scenarios} remains an open problem \cite{yang2026wifo}. 

Due to the inherent graphical structure of MIMO systems \cite{kosasihGraphNeuralNetwork2022a,shen2022graph}, where the transmitted symbols, received signals, and channel coefficients can be naturally represented as nodes and edges in a graph, graph learning-based foundation models \cite{liu2025graph} have emerged as a promising paradigm for MIMO signal processing, {with strong representational ability and scalability for modeling complex interactions and dependencies in MIMO systems.} 
The development of this paradigm requires a meticulous design of the backbone architecture. Although classical GNNs, also referred to as message passing NNs (MPNNs), dominate current applications in wireless communications, their representational limits are quickly reached due to the over-smoothing
bottleneck \cite{liu2025graph}, hindering the performance in large-scale graph pre-training. To overcome this capacity barrier, Transformer-based architectures offer an effective alternative via the global self-attention mechanism. By explicitly incorporating graph topologies as learnable spatial biases within the attention score calculation \cite{ying2021transformers,jumper2021highly}, graph Transformers (GTs) {close the gap left by standard Transformers in graph information encoding}. This endows them with the capability to capture complex, long-range interference topology in wireless networks more effectively than MPNNs \cite{sheng2026graph}. 
Despite these structural advancements, current literature mostly focuses on task-specific GNN/GT detectors, leaving the potential of large-scale pre-trained GTs for signal recovery tasks unexplored.
This naturally raises a research question: \textit{How can we construct a wireless-native graph foundation model capable of scalable and generalized MIMO detection across complex, heterogeneous scenarios?}

In response to this question, this work develops a GFM tailored for large-scale MIMO detection. 
To construct an expressive and scalable backbone, inspired by 
\cite{rampavsek2022recipe}, we propose a physics-informed hybrid backbone architecture that combines MPNN and GT blocks and explicitly encodes physical interference (graph edge features) into the message passing and attention mechanisms, {so that the GFM can capture both local correlations and global dependencies between MIMO data streams}. 
By infusing the statistical principles of expectation propagation (EP)-based MIMO detection \cite{cespedesExpectationPropagationDetection2014} into the edge features and structural design, the GFM inherits the expert knowledge from EP and facilitates accurate interference representation and symbol detection. The GFM undergoes extensive pre-training on a large-scale dataset that encompasses a wide range of system configurations and channel conditions. 
For specific downstream deployment, we introduce parameter-efficient fine-tuning (PEFT) techniques, facilitating rapid and low-overhead adaptation to heterogeneous system regimes widespread in practical large-scale MIMO systems. To alleviate the computational overhead of foundation models, a mixture-of-experts (MoE) structure is embedded to enable input-dependent conditional routing, selectively activating sub-networks to improve inference efficiency. 
Numerical results demonstrate that the proposed GFM detector consistently outperforms both conventional detectors and advanced deep learning baselines across various challenging zero-shot and few-shot conditions. 
{Our evaluations also verify} the efficacy of the overall framework,  spanning pre-training, rapid adaptation, and efficient inference. 
The contributions of this paper are summarized as follows.
\begin{itemize}
  \item \textbf{Physics-Informed GFM Architecture for Large-Scale MIMO Detection:} We develop a novel GFM architecture that integrates MPNNs for local correlation extraction and GT layers for physically grounded global information aggregation via explicit injection of graph edge features as attention biases, {mitigating} the over-smoothing bottleneck. The proposed design also embeds EP detection statistics into the edge features, regularizing the data-driven learning process with domain knowledge.

  \item \textbf{Extensive Pre-Training and Parameter-Efficient Fine-Tuning:} {We establish a two-stage learning workflow to endow the proposed detector with both robust generalization and scenario-aware adaptability.} 
  The GFM is first pre-trained on massive datasets of MIMO transmission to learn generalized representations for MIMO detection that support diverse modulation schemes, antenna configurations, and channel conditions. {Subsequently, PEFT is leveraged to rapidly adapt the pre-trained model to unseen, non-ideal system regimes, e.g., few-bit analog-to-digital converters (ADCs), nonlinear power amplifiers (PAs), or inter-cell interference, with minimal overhead.} 
  
  \item \textbf{Mixture-of-Experts Learning for Efficient Inference:} We embed {an} MoE mechanism into the GFM to improve inference efficiency. This mechanism selectively activates a specialized sub-network (MPNN or GT) within the GFM based on the  characteristics 
  of instantaneous channel realizations,
  {reducing computational cost while maintaining detection accuracy}. 

\end{itemize}

\textit{Notations:} 
For any matrix $\mathbf{A}$, $\mathbf{A}^{\top}$ and $\mathbf{A}^{-1}$ denote the transpose and inverse of $\mathbf{A}$, respectively. 
$\text{diag}(\mathbf{A})$ returns the diagonal vector of a matrix $\mathbf{A}$, $\text{Diag}(\mathbf{a})$ denotes a diagonal matrix with the elements of $\mathbf{a}$ on its main diagonal, and $\lambda_{\max}(\mathbf{A})$ is the largest eigenvalue of $\mathbf{A}$.
$\mathbf{0}$ and $\mathbf{1}$ are the all-zero and all-one vectors, respectively, and $\mathbf{I}$ is an identity matrix.
Also, $\|\cdot\|_1$, $\|\cdot\|_2$, $\|\cdot\|_\infty$, and $\|\cdot\|_F$ denote the {$\ell_1$, $\ell_2$}, infinity, and Frobenius norms, respectively.
$\mathbb{E}[\cdot]$ is the expectation operator, and
$\odot$ and $\oslash$ are the componentwise vector multiplication and division operators, respectively.
Moreover, $\mathbb{R}$ and $\mathbb{C}$ represent the set of real and complex numbers, respectively.
$\mathcal{N}(\mathbf{x}; \boldsymbol{\mu}, \boldsymbol{\Sigma})$ represents the Gaussian distribution of a random vector $\mathbf{x}$ with mean $\boldsymbol{\mu}$ and covariance $\boldsymbol{\Sigma}$.

\section{Problem Formulation and Preliminaries} \label{sec:problem}  
\subsection{System Model}
We consider an uplink multi-user MIMO system, where a base station (BS) equipped with $\nr$ antennas serves $\nt$ single-antenna user equipments (UEs). 
The equivalent real-valued model of this MIMO transmission can be expressed as
\begin{equation} \label{eq:sys_model}
\mathbf{y} = \mathbf{H}\mathbf{x} + \mathbf{n},
\end{equation}
where $\mathbf{y} \in \mathbb{R}^{M \times 1}$ denotes the received signal at the BS with $M=2\nr$, $\mathbf{x} \in \mathbb{R}^{N \times 1}$ denotes the transmitted symbol vector with $N=2\nt$, $\mathbf{H} \in \mathbb{R}^{M \times N}$ is the channel matrix, and $\mathbf{n} \in \mathbb{R}^{M \times 1}$ is the additive white Gaussian noise (AWGN) vector with zero mean and covariance $\sigma^2 \mathbf{I}$. Each transmitted symbol is drawn from a quadrature amplitude modulation (QAM) constellation $\mathcal{S}=\{a_1, a_2, \ldots, a_S\}$ with a cardinality of $|\mathcal{S}|=S$ and a constant average energy $E_{\rm s}=1/2$. The transmission signal-to-noise ratio (SNR) is defined as $\text{SNR}=\mathbb{E}[\|\mathbf{H}\mathbf{x}\|_2^2] / \mathbb{E}[\|\mathbf{n}\|_2^2]$.

The objective of MIMO detection is to recover the transmitted symbol vector $\mathbf{x}$ given $\mathbf{y}$ and the knowledge of $\mathbf{H}$. Assuming a uniform prior, the optimal maximum likelihood (ML) detection criterion is given by 
\CheckRmv{
  \begin{equation} \label{eq:ML}
  \hat{\mathbf{x}}_{\rm ML}=\underset{\mathbf{x} \in \mathcal{S}^{N \times 1}}{\arg \min }\; \|\mathbf{y} - \mathbf{H}\mathbf{x}\|_2^2, 
  \end{equation}
} %
which is an NP-hard problem involving prohibitive complexity when $N$ or $S$ is large.
This computational bottleneck necessitates a judicious performance-complexity trade-off, while practical deployment further demands {adaptability across diverse system configurations and channel conditions}, which existing detectors often fail to meet. {To address this gap,} this work constructs a highly scalable GFM to serve as a {general-purpose backbone for MIMO detection}.

To characterize the MIMO interference, we formulate \eqref{eq:sys_model} over a fully connected graph $\mathcal{G} = (\mathcal{V}, \mathcal{E})$. Here, $\mathcal{V} = \{1, \ldots, N\}$ corresponds to the $N$ unknown symbols in $\mathbf{x}$. Meanwhile, the edge set $\mathcal{E}$ encapsulates the pairwise interactions between these streams, 
which are primarily governed by the inter-stream channel correlations $\mathbf{h}_k^{\top} \mathbf{h}_j$, where $\mathbf{h}_k$ is the $k$-th column of $\mathbf{H}$. By casting the physical transmission into this topological view, the intricate MIMO interference is translated into graph signals, allowing the GFM to extract robust representations.

\CheckRmv{
\begin{table}[t]
  \renewcommand{\arraystretch}{1.3}
  \caption{Typical Downstream System Regimes Causing Model Mismatches}
  \label{tab:downstream}
  \centering
  \setlength\tabcolsep{3pt}
  \begin{tabular}{c l p{4.7cm}}
    \toprule
    \textbf{Variable} & \textbf{System Regime} & \textbf{Mismatch Description} \\
    \midrule
    \multirow{2}{*}{$\mathbf{y}$} & \multirow{2}{*}{Few-bit ADCs \cite{wenBayesOptimalJointChannelandData2016}} & $\bar{\mathbf{y}} = \mathsf{Q}(\mathbf{y})$, where $\mathsf{Q}(\cdot)$ is an ADC quantization function. \\
    \hline
    \multirow{2}{*}{$\mathbf{H}$} & {Oscillator phase} & $\bar{\mathbf{H}} = \boldsymbol{\Theta}_{\rm BS}\mathbf{H}\boldsymbol{\Theta}_{\rm UE}$, where $\boldsymbol{\Theta}$ denotes the  \\
                                  &  {noise \cite{bjornson2014massive}} &   rotation matrix induced by phase drifts. \\
    \hline
    \multirow{2}{*}{$\mathbf{x}$} & \multirow{2}{*}{Nonlinear PAs \cite{goncalves2025optimum}} & $\bar{x} = \mathsf{PA}(\tilde{x})$, where $\tilde{x}\in \mathbb{C}$ is the  complex transmitted symbol. \\
    \hline
    \multirow{2}{*}{$\mathbf{n}$} & Inter-cell & $\bar{\mathbf{n}} \sim \mathcal{N}\big(\mathbf{0}, \rho\mathbf{C}\mathbf{C}^{\top} + \sigma^2\mathbf{I}\big)$ with power  \\
                                  & interference \cite{zhao2025efficient} & $\rho$ and channel $\mathbf{C}$ of non-target UEs. \\ 
    \bottomrule
  \end{tabular}
\end{table}
}

\subsection{Downstream System Regimes} \label{sec:downstream}  
In practice, the ideal transmission model in \eqref{eq:sys_model} is inevitably distorted by various hardware imperfections and environmental non-idealities, leading to significant mismatches that can severely degrade performance. 
Therefore, the pre-trained GFM is expected to be swiftly adaptable to heterogeneous downstream system regimes that encompass such model mismatches via few-shot learning. Typical examples of these system regimes and their corresponding mismatch descriptions are summarized in \tabref{tab:downstream}. 
{In particular, we categorize these examples based on the mismatches and impairments on the four variables within \eqref{eq:sys_model}, guiding the subsequent physical feature design for the MoE router and the evaluations of the proposed PEFT modules against all dimensions of distortion. The specific mismatches are detailed as follows:}

\textbf{Few-Bit ADCs ($\mathbf{y}$):} To reduce the power consumption and hardware cost of massive antenna arrays, low-resolution ADCs are increasingly employed at the BS \cite{wenBayesOptimalJointChannelandData2016}, yielding a quantized received signal $\bar{\mathbf{y}} = \mathsf{Q}(\mathbf{y})$, where the nonlinear quantization function $\mathsf{Q}(\cdot)$ maps the continuous signal to a discrete set. This regime requires the detector to recover signals from severely distorted observations under low-resolution configurations, such as 3-bit quantization.

{\textbf{Oscillator Phase Noise ($\mathbf{H}$):} In large-scale multi-user MIMO systems, maintaining a globally synchronized carrier and mitigating phase drifts are challenging. Independent oscillators at the BS and UEs introduce multiplicative phase noise, perturbing the effective channel to $\bar{\mathbf{H}} = \boldsymbol{\Theta}_{\rm BS}\mathbf{H}\boldsymbol{\Theta}_{\rm UE}$ \cite{bjornson2014massive}.  To align with the real-valued model, both rotation matrices take the form {$\boldsymbol{\Theta}=[\mathbf{C}_{\theta}, -\mathbf{S}_{\theta}; \mathbf{S}_{\theta}, \mathbf{C}_{\theta}]$ with $\mathbf{C}_{\theta}= \text{Diag}(\cos\boldsymbol{\theta})$ and $\mathbf{S}_{\theta}= \text{Diag}(\sin\boldsymbol{\theta})$}. The phase drift vectors $\boldsymbol{\theta}$ at both ends comprise independent elements following $\mathcal{N}(0, \sigma_{\theta}^2)$. Relying on the static $\mathbf{H}$ ignores this unmodeled drift, causing a severe mismatch that compromises conventional detection \cite{pitarokoilis2015uplink}.}

\textbf{Nonlinear PAs ($\mathbf{x}$):} {To maximize power efficiency in the uplink transmission, PAs at UEs often operate near the saturation region \cite{goncalves2025optimum}. For modern solid-state PAs, this predominantly manifests as a nonlinear amplitude distortion of the transmitted symbols $\tilde{x}\in \mathbb{C}$ without significant phase shift, which follows the Rapp model \cite{rapp1991effects}. The distorted symbol $\bar{x}=\mathsf{PA}(\tilde{x})=A(|\tilde{x}|)e^{j \angle (\tilde{x})}$ maintains its original phase $\angle (\tilde{x})$, while its amplitude is compressed as
\CheckRmv{
  \begin{equation}
    A(|\tilde{x}|) = \frac{|\tilde{x}|}{\left(1 + (|\tilde{x}| / A_{\rm sat})^{2p}\right)^{\frac{1}{2p}}},
  \end{equation}
}
where $A_{\rm sat}$ is the saturation level, and $p$ 
is the smoothness factor.} Consequently, the decision boundaries of the detector should be effectively calibrated to account for this distortion.

\textbf{Inter-Cell Interference ($\mathbf{n}$):} In multi-cell dense deployments, the target signal is often corrupted by inter-cell interference from non-target UEs. Instead of being white, interference from $N_{\rm int}$ non-target UEs plus AWGN becomes colored. Specifically, the colored noise follows a Gaussian distribution, denoted as $\mathcal{N} (\mathbf{0}, \rho \mathbf{C}\mathbf{C}^{\top} + \sigma^2 \mathbf{I})$ \cite{zhao2025efficient}, where $\mathbf{C} \in \mathbb{R}^{M \times 2N_{\rm int}}$ denotes the channel matrix of $N_{\rm int}$ non-target UEs, and $\rho$ signifies their power. 
The detector should therefore adapt to handle the non-trivial noise covariance, {enhancing robustness} in interference-limited scenarios.

\subsection{Expectation Propagation-Based Detection} \label{sec:ep}

According to Bayes' theorem, the exact posterior distribution of the transmitted signal $\mathbf{x}$ is given by 
\CheckRmv{
  \begin{equation}
    p(\mathbf{x}|\mathbf{y}) \propto \mathcal{N}(\mathbf{y}; \mathbf{H}\mathbf{x}, \sigma^2\mathbf{I}) \prod_{i=1}^N p(x_i),
  \end{equation}
}
where $p(x_i)$ denotes the discrete prior over the constellation $\mathcal{S}$. Since recovering $\mathbf{x}$ from this exact posterior requires a prohibitively exhaustive search, the EP algorithm \cite{seegerExpectationPropagationExponential2005,cespedesExpectationPropagationDetection2014} is used to iteratively approximate the joint posterior by modeling the discrete priors as independent Gaussian distributions.

Specifically, at the $l$-th iteration ($l=1,2,\ldots,L$), the approximated posterior is formulated as a Gaussian distribution $q_{l}(\mathbf{x}) = \mathcal{N}\big(\mathbf{x}; \boldsymbol{\mu}_{l}, \boldsymbol{\Sigma}_{l}\big)$. Its covariance matrix and mean vector are obtained using an LMMSE estimation step as follows: 
\CheckRmv{
  \begin{subequations} \label{eq:lmmse}%
  \begin{align}
    \boldsymbol{\Sigma}_{l} &= \big(\sigma^{-2}\mathbf{H}^{\top}\mathbf{H} + \boldsymbol{\Lambda}_{l-1}\big)^{-1}, \label{eq:ep_sigma} \\
    \boldsymbol{\mu}_{l} &= \boldsymbol{\Sigma}_{l}\big(\sigma^{-2}\mathbf{H}^{\top}\mathbf{y} + \boldsymbol{\gamma}_{l-1}\big), \label{eq:ep_mu}
  \end{align}
  \end{subequations}
}
where $\boldsymbol{\gamma}_{l-1}$ and $\boldsymbol{\Lambda}_{l-1} = \text{Diag}(\boldsymbol{\lambda}_{l-1})$ {represent the information vector and precision matrix, respectively, provided by} the $(l-1)$-th iteration ($\boldsymbol{\gamma}_0=\mathbf{0}$ and $\boldsymbol{\lambda}_0 = \frac{1}{E_{\rm s}}\mathbf{1}$). 
To progressively refine these parameters, EP computes a cavity distribution $\mathcal{N}(x_i; x_{{\rm e},l,i}, v_{{\rm e},l,i})$ for each variable $x_i$, which effectively isolates the self-feedback information of the current node to avoid loop accumulation \cite{cespedesExpectationPropagationDetection2014,zhou2023graph}. Specifically, the cavity mean, $\mathbf{x}_{{\rm e},l}=[x_{{\rm e},l,1}, \ldots, x_{{\rm e},l,N}]^{\top}$, and variance, $\mathbf{v}_{{\rm e},l}=[v_{{\rm e},l,1}, \ldots, v_{{\rm e},l,N}]^{\top}$, can be computed as
\CheckRmv{
  \begin{subequations} \label{eq:cavity}%
    \begin{align}
      \mathbf{v}_{{\rm e},l} &= \text{diag}({\boldsymbol{\Sigma}_{l}} ) \oslash \left(\mathbf{1}-\text{diag}(\boldsymbol{\Sigma}_{l}) \odot \boldsymbol{\lambda}_{l-1}\right), \\
      \mathbf{x}_{{\rm e},l} &= \mathbf{v}_{{\rm e},l} \odot \left({\boldsymbol{\mu}_{l}}\oslash{\text{diag}(\boldsymbol{\Sigma}_{l})}-\boldsymbol{\gamma}_{l-1}\right).
    \end{align}%
  \end{subequations}
}

Then, the approximated posterior distribution of $x_i$ is computed by incorporating the true prior $p(x_i)$ as $\hat{p}_l(x_i|\mathbf{y})\propto\mathcal{N}(x_i; x_{{\rm e},l,i}, v_{{\rm e},l,i})\cdot p(x_i)$, whose mean $\hat{x}_{l,i}$ and variance $ \hat{v}_{l,i}$ are given by  
\CheckRmv{
  \begin{subequations} \label{eq:post}%
    \begin{align}
      \hat{x}_{l,i} &= \sum_{a_s \in \mathcal{S}} a_s \cdot \hat{p}_l(x_i=a_s|\mathbf{y}), \\
      \hat{v}_{l,i} &= \sum_{a_s \in \mathcal{S}} (a_s - \hat{x}_{l,i})^2 \cdot \hat{p}_l(x_i=a_s|\mathbf{y}).  
    \end{align}
  \end{subequations}
}
These posterior means and variances constitute the vectors $\hat{\mathbf{x}}_l = [\hat{x}_{l,1}, \ldots, \hat{x}_{l,N}]^{\top}$ and $\hat{\mathbf{v}}_l = [\hat{v}_{l,1}, \ldots, \hat{v}_{l,N}]^{\top}$. Finally, the parameters $\boldsymbol{\lambda}_{l}$ and $\boldsymbol{\gamma}_{l}$ are updated by matching the first two moments of the independent Gaussian distributions with the approximated posterior distributions as follows:
\CheckRmv{
  \begin{subequations} \label{eq:moment_matching}%
    \begin{align}
      \boldsymbol{\lambda}_{l} &= \beta \left(\mathbf{1} \oslash \hat{\mathbf{v}}_l -\mathbf{1} \oslash {\mathbf{v}}_{{\rm e},l}\right) +  (1-\beta)\boldsymbol{\lambda}_{l-1}, \\
      \boldsymbol{\gamma}_{l} &= \beta \left(\hat{\mathbf{x}}_l \oslash \hat{\mathbf{v}}_l -{\mathbf{x}}_{{\rm e},l} \oslash {\mathbf{v}}_{{\rm e},l}\right) +  (1-\beta)\boldsymbol{\gamma}_{l-1},
    \end{align}
  \end{subequations}
}
where $\beta\in [0,1]$ is a damping factor to smooth the updates, and the non-negativity of $\boldsymbol{\lambda}_{l}$ is enforced by setting 
$\lambda_{l,i} = \lambda_{l-1,i}$ and $\gamma_{l,i} = \gamma_{l-1,i}$ when $1/\hat{v}_{l,i} - 1 / {v}_{{\rm e},l,i} < 0$. 
After $L$ iterations, the final detection output is given by $\hat{\mathbf{x}}_L$, on which a hard decision is applied.

While EP provides a rigorous inference framework, its independent Gaussian approximation fails to capture complex interference relationships, particularly when multi-stream interference is severe \cite{kosasihGraphNeuralNetwork2022a}. To overcome this bottleneck while preserving its domain insights, our proposed method explicitly embeds intermediate EP statistics as physics-informed graph edge features, regularizing the data-driven mechanisms within the GFM for {robust symbol recovery}.

\section{Proposed Approaches} \label{sec:method}   
In this section, we first present the detailed model architecture of the proposed GFM for MIMO detection. We then elaborate on the learning workflow, which encompasses a {pre-training procedure} to establish a general detection model, followed by PEFT strategies for rapid adaptation to diverse downstream regimes. Finally, we introduce the MoE scheme designed to {reduce computational cost during inference}.

\CheckRmv{
  \begin{figure*}[t]
    \setlength{\abovecaptionskip}{-0.1cm}
		\setlength{\belowcaptionskip}{-0.0cm}
    \centering
    \includegraphics[width=7.0in]{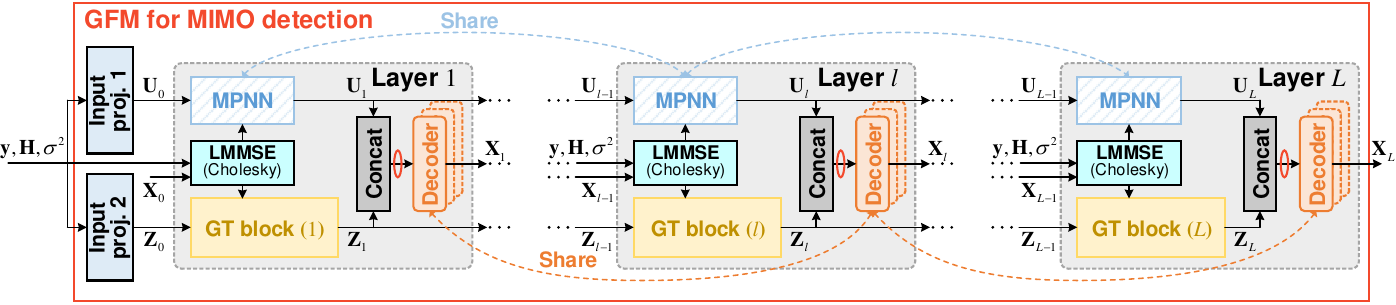}
    \caption{Overall architecture of the proposed GFM for MIMO detection. The MPNN and decoder heads share parameters across layers, while the GT blocks are layer-specific.}
    \label{fig:overall}
  \end{figure*}
}

\subsection{Model Architecture} 
The overall architecture of the proposed GFM is illustrated in \figref{fig:overall}. 
The GFM cascades $L$ processing layers, 
each integrating an LMMSE module inherited from EP to explicitly compute physical statistics, together with data-driven MPNN and GT blocks to capture interference representations, followed by the decoder heads to produce the detection results. 
To clarify the structure, we first formulate the layer-wise operation flow, and subsequently delve into the detailed designs of each constituent module.

\subsubsection{Overall Operational Flow}
To map MIMO detection into a graph learning paradigm, the GFM is established upon carefully designed node features, edge representations, and detection statistics. 
Specifically, the initial node feature $\mathbf{F} \in \mathbb{R}^{N\times 2}$ is constructed based on the graph convolution approximation of $\mathbf{x}$ \cite[Eq.~(23)]{lu2024gcepnet} as 
\CheckRmv{
  \begin{equation}
    \mathbf{F} = \left[\eta\mathbf{H}^{\top}\mathbf{y}, \eta\sigma\mathbf{H}^{\top}\mathbf{1}\right],\; \eta \triangleq 1 / \lambda_{\max}(\mathbf{H}^{\top}\mathbf{H}),
  \end{equation}
} 
representing the initial knowledge of the nodes, where $\sigma\mathbf{H}^{\top}\mathbf{1}$ is an approximation of $\mathbf{H}^{\top}\mathbf{n}$, and $\eta$ is a normalization factor. 
Subsequently, $\mathbf{F}$ is linearly projected to generate the initial embeddings for the MPNN and GT branches:
\CheckRmv{
  \begin{align}
    \mathbf{U}_0 &= \mathsf{Linear}_{\rm U}\left(\mathbf{F}\right)\in \mathbb{R}^{N\times d_{\rm U}}, \\
    \mathbf{Z}_0 &= \mathsf{Linear}_{\rm Z}\left(\mathbf{F}\right)\in \mathbb{R}^{N\times d_{\rm Z}},
  \end{align}
}
where $d_{\rm U}$ and $d_{\rm Z}$ denote the respective latent feature dimensions.
Moreover, the prior detection statistics are initialized as $\mathbf{X}_0\triangleq[\hat{\mathbf{x}}_{0}, \hat{\mathbf{v}}_{0},\boldsymbol{\gamma}_0,\boldsymbol{\lambda}_0]\in \mathbb{R}^{N\times 4}$, where $\boldsymbol{\gamma}_0=\hat{\mathbf{x}}_{0}=\mathbf{0}$, $\hat{\mathbf{v}}_{0}=E_{\rm s}\mathbf{1}$, and $\boldsymbol{\lambda}_0=\mathbf{1}\oslash{\hat{\mathbf{v}}_{0}}$. 
Together, they form the initial state tuple $\{\mathbf{U}_{0}, \mathbf{Z}_{0}, \mathbf{X}_{0}\}$ of the GFM.

At each layer $l=1,\ldots,L$, the state tuple of the GFM, $\{\mathbf{U}_{l-1}, \mathbf{Z}_{l-1}, \mathbf{X}_{l-1}\}$, is iteratively updated by exchanging information across the graph edges. Following the relationships characterized by the graphical model's pair potential \cite[Eq.~(6)]{scottiGraphNeuralNetworks2020}, the physics-informed edge feature tensor at layer $l$, denoted as {$\mathbf{E}_l = \{ \mathbf{e}_{l,jk} \}_{j,k=1}^N \in \mathbb{R}^{N \times N \times 6}$}, is constructed as 
\CheckRmv{
  \begin{equation} \label{eq:edge_feature}
    {\mathbf{e}_{l,jk} = \left[\mathbf{h}_{k}^{\top} \mathbf{h}_{j}, \sigma^{2}, x_{{\rm e},l, j}, x_{{\rm e},l, k}, v_{{\rm e},l, j}, v_{{\rm e},l, k}\right].}
  \end{equation}
}
This feature encapsulates both the static inter-stream correlations and the dynamic node-specific extrinsic statistics iteratively inferred by the EP principle. 
Specifically, the LMMSE module first uses the prior statistics in $\mathbf{X}_{l-1}$, along with the global state $\{\mathbf{y}, \mathbf{H}, \sigma^2\}$, to extract the essential extrinsic information $\mathbf{x}_{{\rm e},l}$ and $\mathbf{v}_{{\rm e},l}$, thereby assembling $\mathbf{E}_l$. 
This feature is then injected into both the MPNN and GT blocks, guiding the MPNN to update the local representation $\mathbf{U}_{l}$, while acting as attention biases for the GT to refine the global representation $\mathbf{Z}_{l}$. Finally, the decoder heads map the concatenated latent features into the updated detection statistics $\mathbf{X}_l$. 
The layer-wise update process can be summarized as:
\CheckRmv{
  \begin{align}
  \mathbf{U}_{l}, \mathbf{Z}_{l}, \mathbf{X}_{l} &= \mathsf{GFM}_{l}(\mathbf{U}_{l-1}, \mathbf{Z}_{l-1}, \mathbf{X}_{l-1}), \\
  \text{computed as} \quad 
  \mathbf{E}_{l} &= \mathsf{LMMSE}(\mathbf{X}_{l-1}, \mathbf{y}, \mathbf{H}, \sigma^2), \\   
  \mathbf{U}_{l} &= \mathsf{MPNN}(\mathbf{U}_{l-1},\mathbf{E}_{l}), \\
  \mathbf{Z}_{l} &=\mathsf{GT}_{l}(\mathbf{Z}_{l-1},\mathbf{E}_{l}), \\
  \mathbf{X}_{l} &=\mathsf{Decoder}\left(\mathsf{Concat}[\mathbf{U}_{l}, \mathbf{Z}_{l}] \right).
\end{align}
}
To balance training stability and representation expressivity, the MPNN and decoder heads are shared across all layers, whereas the GT blocks are layer-specific to {increase expressive power}. 
Guided by this overall workflow, the designs of the LMMSE module, MPNN, GT, and the decoder heads are elaborated in the following parts.

\subsubsection{LMMSE Module} \label{sec:lmmse}
Serving as the physics-driven core inherited from EP, the LMMSE module extracts essential statistics of the interference. Specifically, it utilizes the approximated prior parameters $\boldsymbol{\lambda}_{l-1}$ and $\boldsymbol{\gamma}_{l-1}$ from $\mathbf{X}_{l-1}$ to compute \eqref{eq:lmmse}, obtaining the mean $\boldsymbol{\mu}_{l}$ and covariance $\boldsymbol{\Sigma}_{l}$ of the Gaussian approximation. 
Instead of explicit matrix inversion in \eqref{eq:ep_sigma}, Cholesky decomposition \cite{golubMatrixComputations2013,wuLargeScaleMIMODetection2014} is first calculated as $\sigma^{-2}\mathbf{H}^{\top}\mathbf{H} + \boldsymbol{\Lambda}_{l-1} = \mathbf{L}_l \mathbf{L}_l^{\top}$, where $\mathbf{L}_l$ is a lower triangular matrix. Then,  $\boldsymbol{\Sigma}_{l}$ can be solved via forward and backward substitutions \cite{studerASICImplementationSoftInput2011}.\footnote{This approach avoids the computationally intensive matrix multiplication involved in computing $(\mathbf{L}_l^{\top})^{-1}\mathbf{L}_l^{-1}$.} {While still exhibiting an $\mathcal{O}(N^3)$ complexity, this hardware-friendly approach guarantees numerical stability and reduces implementation overhead for practical deployments.}
Subsequently, the EP cavity computations in \eqref{eq:cavity} are conducted to produce the extrinsic information $\mathbf{X}_{{\rm e},l} = [\mathbf{x}_{{\rm e},l}, \mathbf{v}_{{\rm e},l}]$, which is aggregated to construct the physics-informed feature {$\mathbf{E}_l$} to guide the subsequent data-driven modules.

\CheckRmv{
  \begin{figure}[t]
    \setlength{\abovecaptionskip}{-0.1cm}
		\setlength{\belowcaptionskip}{-0.0cm}
    \centering
    \includegraphics[width=3.1in]{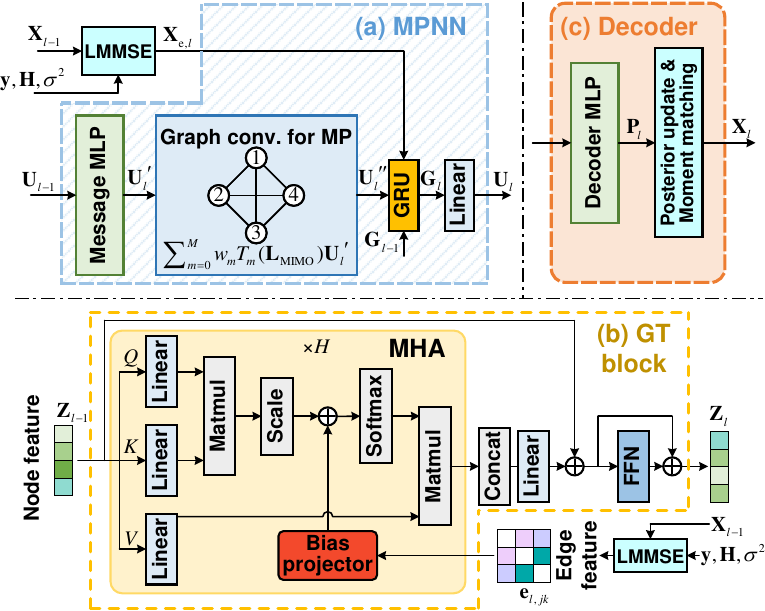}
    \caption{Details of the components within the GFM. (a) Spectral convolution-enhanced MPNN. (b) GT with a bias projector. (c) Decoder head for posterior distribution approximation.}
    \label{fig:model}
  \end{figure}
}

\subsubsection{Spectral Convolution-Enhanced MPNN} 
To capture local interference patterns within the MIMO graph, we employ an MPNN inspired by \cite{lu2024gcepnet}, as illustrated in \figref{fig:model}(a). 
Traditional message passing often struggles with the dense connectivity in large-scale MIMO scenarios. In contrast, our approach utilizes Chebyshev polynomial-based spectral graph filters operating on the normalized MIMO Laplacian, $\mathbf{L}_{\rm MIMO} \triangleq \mathbf{I} - \eta \mathbf{H}^{\top} \mathbf{H}$, explicitly characterizing the spatial correlations and circumventing the bottleneck in densely connected graphs with $N(N-1)$ message computations.

To {improve the generalizability} of the spectral graph filters across dynamic wireless environments, we use an instance-adaptive attention mechanism, denoted as $\mathsf{ATT}(\cdot)$. This mechanism generates dynamic, channel-aware filter coefficients $[\mathsf{ATT}(\mathbf{F})]_c$ based on the initial node feature $\mathbf{F}$, {tailoring} {the Chebyshev polynomials $T_c(\mathbf{L}_{\rm MIMO})$ with order $c=1,2,\ldots,C$} to the current interference geometry.\footnote{The Chebyshev polynomials are given by $T_c(x) = 2xT_{c-1}(x) - T_{c-2}(x)$ with $T_0(x) = 1$ and $T_1(x) = x$.}
The update rule combines this adaptive spectral filtering with a gated recurrent unit (GRU), given by:
\CheckRmv{
  \begin{align}
    \tilde{\mathbf{U}}_l &= \sum_{c=1}^C [\mathsf{ATT}(\mathbf{F})]_c {T}_c (\mathbf{L}_{\rm MIMO})\mathsf{MLP}_{\rm M}(\mathbf{U}_{l-1}), \\
    \mathbf{G}_l  &= \mathsf{GRU}\left(\mathsf{Concat}[\tilde{\mathbf{U}}_l,\mathbf{X}_{{\rm e}, l}], \mathbf{G}_{l-1}\right),\\
    \mathbf{U}_l &= \mathsf{Linear}_{\rm G}(\mathbf{G}_{l}).
  \end{align}
} 
Here, the message multi-layer perceptron (MLP), $\mathsf{MLP}_{\rm M}(\cdot)$, acts as a local feature projector, while the GRU's hidden state $\mathbf{G}_l$ encodes the physics-informed extrinsic statistics $\mathbf{X}_{{\rm e}, l}$ from the LMMSE module. This integration enables the MPNN to effectively learn expressive local representations tailored to the underlying interference structures. 

\subsubsection{Graph Transformer} 
Complementary to the local feature extraction of MPNN, the GT provides global modeling capacity to capture long-range dependencies. 
While vanilla multi-head attention (MHA) \cite{vaswani2017attention} relies heavily on latent semantic similarities among node queries and keys, we explicitly inject the aforementioned edge feature $\mathbf{E}_l$ into the attention mechanism \cite{ying2021transformers,jumper2021highly} to align with the MIMO interference topology, as shown in \figref{fig:model}(b). Specifically, each feature vector $\mathbf{e}_{l,jk}$ constructed in \eqref{eq:edge_feature} is fed into a shallow projector $\phi_{{\rm B},l}(\cdot)$ to yield an $H$-dimensional bias vector, where $H$ denotes the number of parallel attention heads. For the $h$-th head at layer $l$, the pre-Softmax attention score is formulated as
\CheckRmv{
  \begin{equation}
    A_{l,jk}^{(h)} = \frac{\big(\mathbf{z}_{l-1,j}\mathbf{W}^{(h)}_{\rm Q}\big) \big(\mathbf{z}_{l-1,k}\mathbf{W}^{(h)}_{\rm K}\big)^{\top}}{\sqrt{d}} + [\phi_{{\rm B},l}(\mathbf{e}_{l,jk})]_h,
  \end{equation}
}
where {$\mathbf{z}_{l-1,j}\in \mathbb{R}^{1 \times d_{\rm Z}}$} is the $j$-th row of $\mathbf{Z}_{l-1}$, {$\mathbf{W}^{(h)}_{\rm Q}\in \mathbb{R}^{d_{\rm Z} \times d}$ and $\mathbf{W}^{(h)}_{\rm K}\in \mathbb{R}^{d_{\rm Z} \times d}$} are learnable query and key projection matrices, respectively, and $d=d_{\rm Z}/H$ represents the feature dimension per head. By integrating the head-specific bias term $[\phi_{{\rm B},l}(\mathbf{e}_{l,jk})]_h$, certain heads are encouraged to concentrate on strong interference streams, while others capture global correlations.

Subsequently, the attention-weighted representations from all $H$ heads are concatenated, linearly transformed using the value and output projection matrices ({$\mathbf{W}^{(h)}_{\rm V}\in \mathbb{R}^{d_{\rm Z} \times d}$} and $\mathbf{W}_{\rm O}\in \mathbb{R}^{d_{\rm Z} \times d_{\rm Z}}$), and passed through a feed-forward network (FFN) with residual connections as follows:
\CheckRmv{
  \begin{align}
    \mathbf{o}_{l,j} &= \mathsf{Concat}_{h=1}^H \bigg[\sum_{k=1}^N \mathsf{Softmax}_{k} \big(A_{l,jk}^{(h)}\big) \nonumber \\
    &\quad \quad\quad\quad\quad\quad\quad\quad\times \big(\mathbf{z}_{l-1,k}\mathbf{W}^{(h)}_{\rm V}\big) \bigg] \mathbf{W}_{\rm O}, \\
    \mathbf{z}_{l,j} &= \mathsf{FFN}(\mathbf{o}_{l,j})+ \mathbf{z}_{l -1,j}.
  \end{align}
}
This process produces the updated global feature $\mathbf{Z}_l=[\mathbf{z}_{l,1}^{\top}, \ldots, \mathbf{z}_{l,N}^{\top}]^{\top}$, which is then passed to the decoder heads.

\subsubsection{Decoder Heads}  
The decoder heads project the concatenated local and global features into the symbol space, deriving the updated detection statistics $\mathbf{X}_l$. Considering the structural diversity of different modulation schemes, we employ a bank of modulation-specific decoder heads. Specifically, a three-layer MLP dedicated to the target modulation, $\mathsf{MLP}_{{\rm D},\mathcal{S}}(\cdot)$, is applied to the unified representations to produce an enhanced posterior probability estimation $\mathbf{P}_l \in \mathbb{R}^{N \times S}$ as
\CheckRmv{
  \begin{equation}
    \mathbf{P}_l = \mathsf{MLP}_{{\rm D},\mathcal{S}}\left(\mathsf{Concat}[\mathbf{U}_{l}, \mathbf{Z}_{l}]\right),
  \end{equation}
}
where 
the $i$-th row of $\mathbf{P}_l$ represents the estimated probabilities of $x_i$ over each constellation point. 
This architecture allows the GFM to maintain a shared backbone for high-level interference modeling while utilizing specialized heads for precise symbol-level decisions.
This enhanced posterior estimation improves over the standard EP's independent Gaussian approximation introduced in Section~\ref{sec:ep}. 
Given $\mathbf{P}_l$, the posterior mean $\hat{\mathbf{x}}_l$ and variance $\hat{\mathbf{v}}_l$ can be computed following \eqref{eq:post}. Subsequently, moment matching with the damping factor empirically set to $\beta=0.2$ \cite{zhou2023graph} is performed to update $\boldsymbol{\lambda}_{l}$ and $\boldsymbol{\gamma}_{l}$ per \eqref{eq:moment_matching}, which are then concatenated with the posterior moments to assemble $\mathbf{X}_l$ for the next layer.

\subsection{Model {Pre-Training}} \label{sec:pretraining} 
To endow the GFM with {generalizability across different system configurations}, we adopt a {pre-training strategy} that encompasses a wide range of system parameters and channel conditions.
The pre-training dataset, $\mathcal{D}_{\rm pre}$, can be synthesized to cover a broad spectrum of these configurations, which can be categorized into the following three dimensions:
\begin{itemize}
    \item \textbf{Geometric and Array Configurations:} To ensure scalability across different BS and UE scales, the training phase includes various antenna counts ({$\nr$, $\nt$}) and array geometries, e.g.,  uniform linear array (ULA) and uniform planar array (UPA). This enables the GFM to learn universal spatial correlation patterns that are independent of specific array dimensions.
    \item \textbf{Channel Environments and Propagation:} The pre-training encompasses diverse carrier frequencies and channel scenarios, e.g., 3rd generation partnership project (3GPP) technical report (TR) 38.901 urban macrocell (UMa) and urban microcell (UMi) \cite{3gpp38901}. Moreover, by varying the BS-UE distances and scattering environments, the model captures diverse propagation characteristics.
    \item \textbf{Signal and Modulation Diversity:} The GFM is exposed to diverse signal configurations, including varied modulation schemes (16-, 64-, and 256-QAM), and SNR levels. During pre-training, the modulation-specific decoder head is activated for each sample, enabling the model to map shared representations to modulation-specific posteriors.
\end{itemize}

Each training sample is characterized by a tuple $\{\mathbf{H}, \mathbf{x}, \mathbf{y}, \sigma\}$. Here, $\mathbf{H}$ is randomly drawn from the pre-training channel dataset and normalized for unit average power as $\|\mathbf{H}\|_F^2=\nr\nt$. The transmitted symbol vector $\mathbf{x}$ is generated by randomly selecting symbols from the constellation set assigned to the current sample, 
while the received signal $\mathbf{y}$ is computed using \eqref{eq:sys_model} with AWGN. The noise variance $\sigma^2$ is determined by the SNR level that varies across samples.

To {ensure robustness} against practical impairments, imperfect CSI is simulated during pre-training by introducing estimation errors to the channel matrix, modeled as $\hat{\mathbf{H}} = \mathbf{H} + \mathbf{E}$, where $\mathbf{E}$ is a random error matrix with independent Gaussian entries of zero mean and variance $\sigma_{\rm CSI}^2$. Varying CSI error levels are {incorporated into} the training data to ensure the GFM can handle different degrees of channel uncertainty in real-world deployments.
  
To {accommodate} varying real-valued stream dimensions $N$ and constellation sets $\mathcal{S}$, the overall pre-training loss is formulated as the expectation of the empirical cross-entropy over the training data distribution:
\CheckRmv{
  \begin{equation}  \label{eq:pre_loss}
    \mathcal{L}_{\rm pre} = \mathbb{E}_{\{\mathbf{H}, \mathbf{x}, \mathbf{y}, \sigma, N, \mathcal{S}\} \sim\mathcal{D}_{\rm pre}} \Big[ -\frac{1}{N} \sum_{i=1}^{N} \sum_{a_s \in \mathcal{S}} \mathbb{I}_{x_i=a_s} \log {P}_{L, is} \Big],
  \end{equation}
}
where $\mathbb{I}_{(\cdot)}$ is the indicator function representing the true symbol label, and $P_{L, is}$ is the $(i,s)$-th element of the predicted posterior $\mathbf{P}_L$ produced by the activated decoder head $\mathsf{MLP}_{{\rm D},\mathcal{S}}$, denoting the GFM's predicted probability of $x_i=a_s$.

\CheckRmv{
  \begin{figure}[t]
    \centering
    \includegraphics[width=3.1in]{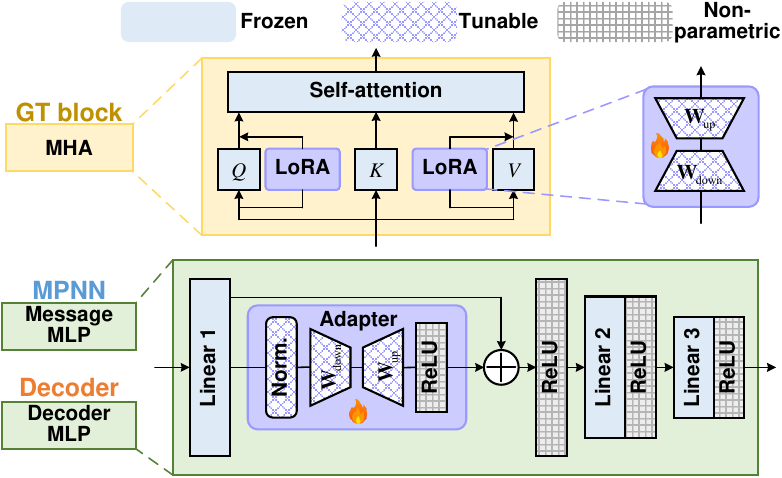}
    \caption{PEFT structures of different components within the GFM.}
    \label{fig:peft}
  \end{figure}
}

\subsection{Parameter-Efficient {Fine-Tuning}}  
To address the inevitable model mismatches in downstream system regimes described in Section~\ref{sec:downstream}, we adapt the pre-trained GFM via lightweight fine-tuning to construct scenario-specific detectors. To ensure rapid adaptation with minimal overhead, we employ PEFT techniques, keeping the majority of the model parameters frozen. Next, we introduce the PEFT strategies tailored to each component within the GFM, as depicted in \figref{fig:peft}.

For the MPNN and the decoder heads, we insert Adapter blocks \cite{gui2024g} into their respective MLPs, i.e., $\mathsf{MLP}_{\rm M}$ and $\mathsf{MLP}_{{\rm D},\mathcal{S}}$.\footnote{Similar to the MPNN and decoder heads, the inserted Adapter blocks are also shared across layers.} From a signal processing perspective, these adapters serve as learnable ``local compensators'' that calibrate the message-passing kernels and decision boundaries to accommodate scenario-specific interference geometries and nonlinear distortions. Specifically, for an input feature $\mathbf{R}$, the Adapter computation is formulated as:
\CheckRmv{
  \begin{equation}
    \mathbf{R}^{\prime} =\mathbf{R} + \mathsf{ReLU}\left(\mathsf{LN}(\mathbf{R})\mathbf{W}_{\rm down}^{(\rm A)}\mathbf{W}_{\rm up}^{(\rm A)}\right),
  \end{equation}
}
where $\mathsf{LN}(\cdot)$ denotes layer normalization, while $\mathbf{W}_{\rm down}^{(\rm A)}$ and $\mathbf{W}_{\rm up}^{(\rm A)}$ are the trainable down- and up-projection matrices constructing a bottleneck structure with dimension $r_{\rm A}$.

Complementary to local adjustment, we utilize a low-rank adaptation (LoRA) approach \cite{hu2022lora} for each GT block to recalibrate the global interference modeling. Since the MHA mechanism relies on learnable query and value projection matrices to capture dependencies across different user streams, LoRA injects trainable low-rank decomposition matrices in parallel to these pre-trained weights:
\CheckRmv{
  \begin{equation}
    \mathbf{Z}^{\prime} = \mathbf{Z}\Big(\mathbf{W}_{\ast} + \frac{\alpha}{r_{\rm L}}\mathbf{W}_{\rm down}^{(\rm L)}\mathbf{W}_{\rm up}^{(\rm L)}\Big),\; \mathbf{W}_{\ast} \in \{\mathbf{W}_{\rm Q}, \mathbf{W}_{\rm V} \},
  \end{equation}
}
where $\mathbf{W}_{\ast}$ denotes the frozen pre-trained projection matrix, $\mathbf{W}_{\rm down}^{(\rm L)}\in \mathbb{R}^{d  \times r_{\rm L}}$ and $\mathbf{W}_{\rm up}^{(\rm L)}\in \mathbb{R}^{r_{\rm L} \times d}$ are the trainable low-rank matrices with $r_{\rm L} \ll d$, and $\alpha \geq r_{\rm L}$ is a scaling factor controlling the adaptation strength. 
{This approach allows the GFM to adapt its attention focus to specific interference conditions with minimal fine-tuning samples.}

By updating only these lightweight parameters, the GFM effectively learns the necessary shift and scale of the latent representations to address downstream mismatches, while preventing catastrophic forgetting of the general detection priors. 
The fine-tuning stage utilizes a limited number of samples, $\mathcal{D}_{\rm FT}$, to align the pre-trained GFM with the  downstream regime. 
The loss function for this stage can be expressed as:
\CheckRmv{
  \begin{equation}
    \mathcal{L}_{\rm FT} = \mathbb{E} \Big[ - \frac{1}{N_{\rm FT}} \sum_{i=1}^{N_{\rm FT}} \sum_{a_s \in \mathcal{S}_{\rm FT}} \mathbb{I}_{x_i=a_s} \log {P}_{L, is}(\boldsymbol{\Theta}_{\rm FT} | \boldsymbol{\Theta}_{\rm Base}) \Big],
  \end{equation}
}
where the expectation is taken over $\mathcal{D}_{\rm FT}$, and $N_{\rm FT}$ and $\mathcal{S}_{\rm FT}$ denote the stream dimension and constellation set for the target downstream regime, respectively, while $\boldsymbol{\Theta}_{\rm FT}$ and $\boldsymbol{\Theta}_{\rm Base}$ represent the trainable adaptation parameters and the frozen pre-trained backbone.
Unlike the pre-training phase, the gradient flow here is restricted to the PEFT modules. This PEFT design enables rapid adaptation with minimal computational and data requirements, facilitating deployment in dynamic wireless environments. 

\CheckRmv{
  \begin{figure}[t]
    \centering
    \includegraphics[width=3.3in]{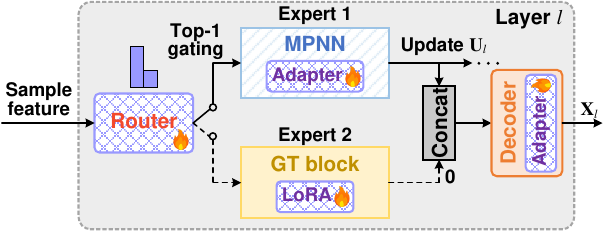}
    \caption{Illustration of the MoE structure in each layer of the GFM.}
    \label{fig:moe}
  \end{figure}
}

\CheckRmv{
  \begin{table}[t]
    \centering
    \begin{threeparttable}
    \caption{Router Architecture}
    \label{tab:router}
    \begin{tabular}{ccc}
      \toprule
      \multicolumn{3}{l}{\textbf{Input}: The physical feature $\mathbf{g}=\psi(\mathbf{y},\mathbf{H},\sigma^2)$} \\
      \midrule
      \textbf{FC layers} & \textbf{Input dim./Output dim.} & \textbf{Activation} \\
      1 & $\text{dim}(\mathbf{g})$/32 & BN+ReLU \\
      2 & 32/16 & BN+ReLU \\
      3 & 16/2 & BN+Softmax \\
      \midrule
      \multicolumn{3}{l}{\textbf{Output}: A binary expert-selection signal} \\
      \bottomrule
    \end{tabular}
    \begin{tablenotes}
      \footnotesize
      \item FC: Fully connected. BN: Batch normalization. 
    \end{tablenotes}
    \end{threeparttable}
  \end{table}
}

\subsection{{Mixture-of-Experts Learning for Inference Efficiency}} 
To balance performance and efficiency, we integrate an MoE mechanism \cite{fedus2022switch} into the GFM architecture. The primary motivation is to mitigate the inference overhead of foundation models, which typically activate their entire parameter set regardless of the task difficulty. Since each GFM layer consists of parallel MPNN and GT branches, these two components can be treated as specialized experts with complementary receptive fields. Although using both experts concurrently {improves generalization}, activating a specific expert is often sufficient to handle certain instances. Therefore, we integrate an MoE structure with top-1 gating within each layer of the model to dynamically select the most relevant expert for each input, {reducing computational overhead and energy consumption} during inference while maintaining detection performance.

As illustrated in \figref{fig:moe}, this MoE mechanism is leveraged during the fine-tuning stage. By combining it with PEFT techniques, we develop a mixture-of-adaptations approach \cite{wang2022adamix,liu2024moe} to facilitate efficient deployment across downstream regimes. To achieve this, a lightweight router is designed to evaluate the physical features of the current input, $\mathbf{g}=\psi(\mathbf{y},\mathbf{H},\sigma^2)$, and output a binary selection signal. 
As detailed in \tabref{tab:router}, the router consists of three fully connected layers followed by a Softmax activation, where the resulting probabilities are converted into a binary decision via an argmax operation. {Based on this signal, the MoE gates the entire computation of the experts, allowing only the selected branch (MPNN or GT) to process the input. To preserve the pre-trained dual-branch template} for the subsequent decoder projection, zero tensors are padded in place of the bypassed expert's output.

This dynamic routing policy adapts to varying channel conditions. In regimes with minimal interference, the router tends to prioritize the simple MPNN expert. Conversely, under severe nonlinear distortions or dense interference environments, it can route the signal to the more expressive GT expert. During fine-tuning, the router is jointly learned with the PEFT parameters to establish an appropriate expert selection strategy for different input characteristics. During inference, this input-dependent selection {reduces the average computational complexity with minimal performance degradation}.

To empower the router to accurately perceive diverse transmission conditions, we empirically choose a set of features that are indicative of essential channel and interference characteristics to construct vector $\mathbf{g}$. 
Specifically, it is defined as
\CheckRmv{
  \begin{equation}
    \mathbf{g}= [g_{{\rm int}}, g_{{\rm dis}}, g_{{\rm peak}}, g_{{\rm SNR}}]^{\top}.  
  \end{equation}
} 
Here, $g_{{\rm int}} = \frac{\| \mathbf{H}^{\top}\mathbf{H}\|_F^2 - \| \text{diag}(\mathbf{H}^{\top}\mathbf{H}) \|_2^2}{\| \text{diag}(\mathbf{H}^{\top}\mathbf{H}) \|_2^2}$ denotes the inter-stream interference ratio, quantifying the orthogonality of the channel matrix; $g_{{\rm dis}} = {\|\mathbf{y}\|_1} / {(\sqrt{M} \|\mathbf{y}\|_2)}$ captures the nonlinear distortion induced by few-bit ADCs; {$g_{{\rm peak}} = {\sqrt{M} \|\mathbf{y}\|_\infty} / {\|\mathbf{y}\|_2}$ measures the normalized peak magnitude, reflecting the amplitude compression caused by PA saturation;} $g_{{\rm SNR}} = {\|\mathbf{y}\|_2} / {(\sqrt{M} \sigma)}$ {reflects the empirical SNR, which is further coupled with phase noise drifts.} 
These features provide multifaceted insights into the characteristics of each detection instance, enabling the router to make informed expert selections.

\section{Numerical Results}  \label{sec:results} 

\subsection{Experimental Details} \label{sec:simu_setup}

\subsubsection{Simulation Settings}
The detailed settings regarding the channel scenarios,  foundation datasets construction, and downstream system regimes are described as follows.

\textbf{Channel Scenarios and Foundation Datasets:} 
To simulate practical massive MIMO uplink transmissions, we generate {diverse channel datasets} using the QuaDRiGa toolbox \cite{jaeckelQuaDRiGa3DMultiCell2014} compliant with the 3GPP TR 38.901 standard \cite{3gpp38901}. As outlined in Section~\ref{sec:pretraining}, a variety of system configurations and scenarios are included to learn a general GFM for massive MIMO detection. Specifically, we simulate non-line-of-sight (NLoS) environments where a BS, placed at a height of 25~m, utilizes various antenna array geometries, including ULA and 3GPP-3D planar configurations. The BS serves single-antenna UEs with a height of 1.5~m uniformly distributed within a distance of 50~m to 300~m, dispersed across either a $120^\circ$ sector or a 100~m radius circle.

The detailed configurations for the pre-training and general
testing channel datasets are summarized in \tabref{tab:datasets}. For pre-training, channel sampling is executed at center frequencies $f_{\rm c} \in \{2.6, 5.9\}$ GHz. Frequency-domain transformation is conducted over 128 effective subcarriers within a 10 MHz bandwidth. Each pre-training dataset from D1 to D5 encompasses six representative MIMO dimensions, i.e., $\nr \times \nt\in\{\text{512}\times \text{128}, \text{256}\times \text{128}, \text{128}\times \text{64}, \text{128}\times \text{32}, \text{64}\times \text{32}, \text{64}\times \text{16}\}$. This yields 38,400 channel samples (6 setups $\times$ 50 user drops $\times$ 128 subcarriers) per dataset, which are partitioned for training and validation at a 24:1 ratio. 
{During testing, the robustness of the GFM is examined over three subsets with different configurations and spatially non-overlapping user drops compared to pre-training datasets, preventing data leakage.} The in-distribution (ID) set D6 comprises 12,800 samples under the \Times{512}{128} and \Times{256}{128} MIMO setups, featuring an unseen combination of system parameters compared to the pre-training datasets. Zero-shot generalization is examined via two OOD sets of 6,400 samples each: {D7 introduces a rural macrocell (RMa) environment under a \Times{96}{32} MIMO setup, and D8 further shifts to an upper-6 GHz (U6G) band with a \Times{1024}{128} XL-MIMO configuration.}

\CheckRmv{
  \begin{table*}[t]
  \centering
  \caption{Overview of the Foundation Channel Datasets}
  \label{tab:datasets}
  \begin{threeparttable}
    \begin{tabular}{ccccccc}
      \toprule
      Set Type & Index & $f_c$ (GHz) & Antenna Array & Channel Scenario & UE Distribution & Num. of Samples \\
      \midrule
      \multirow{5}{*}{Pre-training Set} & D1 & 2.6, 5.9 & 3GPP-3D & UMa & Sector & 38,400 \\
      & D2 & 2.6, 5.9 & ULA & UMa & Sector & 38,400 \\
      & D3 & 2.6, 5.9 & ULA & UMa & Circle & 38,400 \\
      & D4 & 2.6, 5.9 & 3GPP-3D & UMi & Sector & 38,400 \\
      & D5 & 2.6, 5.9 & ULA & UMi & Sector & 38,400 \\
      \midrule
      \multirow{3}{*}{\shortstack{General Testing Set}} & D6 (ID) & 2.6, 5.9 & ULA & UMi & Circle & 12,800 \\
      & D7 (OOD) & 5.9 & 3GPP-3D & RMa & Sector & 6,400 \\
      & D8 (OOD) & 6.8 (U6G) & 3GPP-3D & {RMa} & Sector & 6,400 \\
      \bottomrule
    \end{tabular}
  \end{threeparttable}
\end{table*}
}

\textbf{Signal and Error Configurations:} 
To enrich signal diversity during training data generation, we randomly apply  modulation schemes ranging from 16-QAM to 256-QAM alongside SNRs varied across {{[4, 36]}}~dB. To emulate practical channel acquisition imperfections, random CSI estimation errors are injected into 50\% of the training samples. The error variance $\sigma_{\rm CSI}^2$ is selected to maintain a CSI normalized mean squared error (NMSE) between {{[$-$27, $-$10]}}~dB, {encouraging the GFM to learn robust detection mappings under channel uncertainties}.

\textbf{Downstream Regime Configurations:}
Beyond the foundation datasets, we specialize the pre-trained GFM for the representative downstream regimes discussed in Section~\ref{sec:downstream} to further assess adaptability. 
Specifically, for the \textit{few-bit ADCs} regime, a standard mid-rise uniform quantizer \cite{wenBayesOptimalJointChannelandData2016} is employed with a 3-bit resolution to emulate significant quantization errors. 
{For \textit{oscillator phase noise}, the standard deviations of the phase drifts for the UEs and the BS antennas are set to $\sigma_{\theta,\rm UE} = \sigma_{\theta,\rm BS} = 4^\circ $ ($\approx 0.07$ rad) to reflect a typical impairment level \cite{pitarokoilis2015uplink}.}
{For \textit{nonlinear PAs}, UE-side distortion is characterized via the Rapp model with a saturation level $A_{\rm sat} = 1$ and a smoothness factor $p=2$ according to \cite{goncalves2025optimum}, alongside 16-QAM to trigger amplitude-dependent compression.}
Finally, regarding \textit{inter-cell interference}, interference intensity is quantified by the interference-plus-noise-to-thermal ratio, fixed at $10 \log_{10} ((\rho + \sigma^2) / \sigma^2) = 10$~dB to emulate practical scenarios \cite{zhao2025efficient}. 
For each regime, a dataset of 11,520 samples is constructed (5,120 for fine-tuning, 6,400 for testing), facilitating few-shot performance validation under fine-tuning sample sizes ranging from 128 to 5,120.

\subsubsection{Model Hyperparameters and Training Details} \label{sec:hyperparameter}
Next, we introduce the GFM hyperparameters and training details during the pre-training and fine-tuning phases.

\textbf{Implementation and Pre-training Setup}: All models were implemented in PyTorch and trained on an NVIDIA H100 GPU. The GFM's backbone cascades $L=5$ layers, where each GT block has  a hidden dimension of $d_{\rm Z}=256$ and $H=16$ attention heads. Within the MPNN and the decoder heads, the three-layer MLPs' hidden dimensions are set to $d_{\rm h1}=128$ and $d_{\rm h2}=64$, with a message size of $d_{\rm U}=16$. The spectral convolution order of the MPNN is fixed at $C=3$. 
These hyperparameter selections result in a compact model with 2.19M parameters, striking a balance between model expressiveness, memory cost, and computational efficiency.

The GFM is pre-trained for 200 epochs using the Adam optimizer with an initial learning rate of $7\times 10^{-4}$ and a batch size of 64. To ensure stability, the learning rate is halved if the validation loss plateaus for 10 consecutive epochs, and the minimum learning rate is set to $1\times 10^{-6}$. 

\textbf{Fine-tuning Details}: The fine-tuning phase for each downstream regime is conducted for 50 epochs unless noted otherwise. For the PEFT structures, the Adapter bottleneck dimension is set to $r_{\rm A}=4$, and the LoRA employs a rank of $r_{\rm L}=4$ with a scaling factor of $\alpha=8$.  Given the limited number of trainable parameters, the learning rates for the LoRA and Adapter blocks are elevated to $1\times 10^{-3}$ and $5\times 10^{-3}$, respectively. Furthermore, early stopping with a patience of 10 epochs is enforced to
mitigate overfitting. 

\subsubsection{Baselines}
We benchmark the GFM against a representative series of traditional and learning-based MIMO detectors. Although the NN baselines have smaller parameter scales, they represent highly competitive state-of-the-art references: 
\begin{itemize}
  \item \textbf{LMMSE \cite{yangFiftyYearsMIMO2015}:} The standard linear estimator implemented via the low-complexity Cholesky decomposition-based inverse described in Section~\ref{sec:lmmse}, without iterative refinement.
  \item \textbf{EP \cite{cespedesExpectationPropagationDetection2014}:} The standard EP-based MIMO detector detailed in Section~\ref{sec:ep}, with $L=5$, also adopts the Cholesky decomposition-based inverse.
  \item \textbf{GEPNet \cite{kosasihGraphNeuralNetwork2022a}:} A state-of-the-art learning-based MIMO detector that enhances EP using a fully connected MPNN. We adopt its publicly available implementation\footnote{Available at \url{https://github.com/GNN-based-MIMO-Detection/GNN-based-MIMO-Detection}.} and train it under the same settings as the GFM.
  \item \textbf{EP-GT:} A strong attention-based baseline that takes the EP's estimation as input and refines it using a GT \cite{ying2021transformers}. For a fair comparison, the GT incorporates $L=5$ layers with identical architecture, hyperparameters, and training strategies as the proposed GFM.
\end{itemize}

\subsection{In-Distribution Performance} %
\CheckRmv{
  \begin{figure}[t]
    \centering
      \includegraphics[width=3.45in]{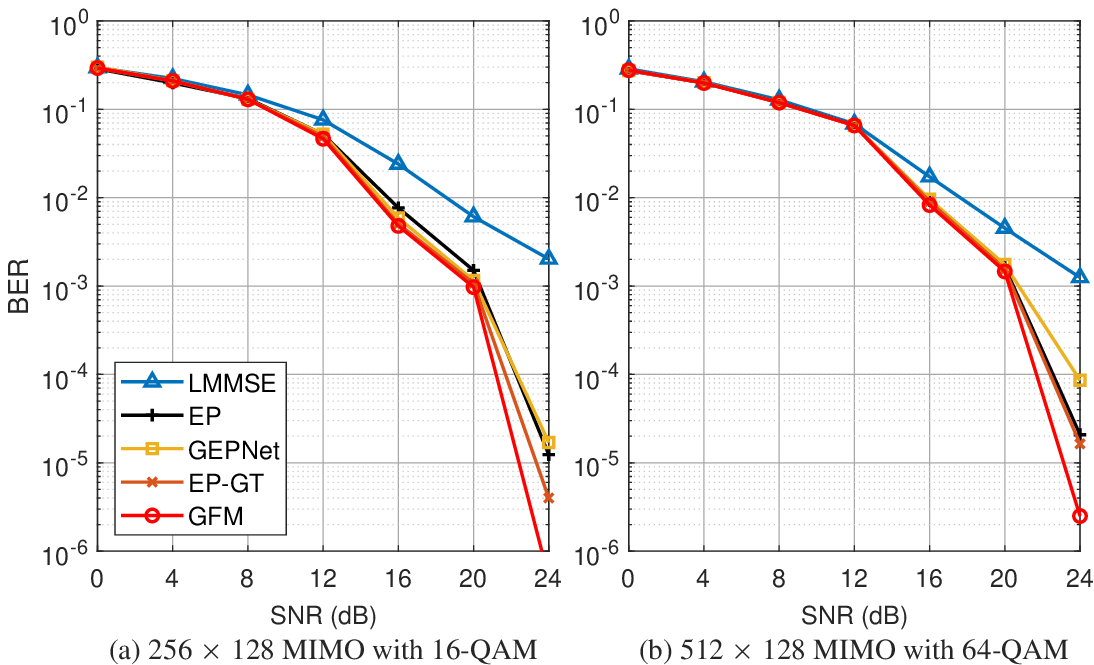}
    \caption{BER performance on the ID testing dataset D6.}
    \label{fig:id}
  \end{figure}
}

\CheckRmv{
  \begin{figure}[t]
    \centering
    \includegraphics[width=3.45in]{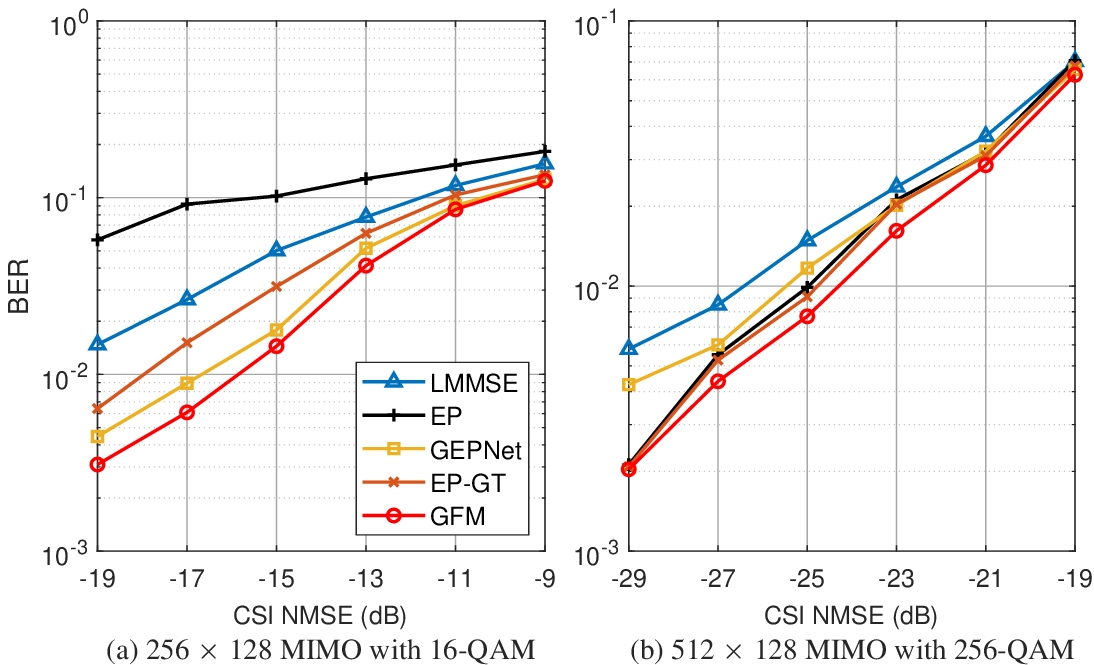}
    \caption{BER performance under imperfect CSI on the ID testing dataset D6, where (a) is at an SNR of 24~dB, and (b) is at an SNR of 28~dB.}
    \label{fig:id_csi}
  \end{figure}
}

We first evaluate the GFM's performance on the ID testing dataset D6. 
The performance is measured in terms of bit error rate (BER) across varying SNR levels and presented in \figref{fig:id}. {Throughout our evaluations, the BER is accumulated until a maximum of $10^7$ bits are transmitted.}
The results demonstrate that the GFM achieves {improved} BER performance across all SNR levels, outperforming traditional methods and learning-based detectors. Specifically, as shown in \figref{fig:id}(a), the GFM achieves an approximate 0.5~dB gain over the best baseline, EP-GT, at a BER of $10^{-4}$ for the \Times{256}{128} MIMO with 16-QAM modulation. Under the \Times{512}{128} MIMO with the more challenging 64-QAM modulation depicted in \figref{fig:id}(b), the GFM maintains its performance advantage compared to all baselines. This indicates that the GFM has effectively learned robust detection mappings during pre-training.

We next investigate the GFM's robustness under imperfect CSI in \figref{fig:id_csi}, where the test NMSE ranges (extending down to $-$29~dB and up to $-$9~dB) are set wider than the pre-training interval of [$-$27, $-$10]~dB to evaluate the extrapolation ability. Specifically, \figref{fig:id_csi}(a) evaluates the \Times{256}{128} MIMO setup with 16-QAM, while \figref{fig:id_csi}(b) considers the \Times{512}{128} MIMO setup with 256-QAM. 
As observed, the GFM sustains a {clear performance edge} over both conventional and learning-based baselines in both scenarios. {For instance, to achieve a target BER of $10^{-2}$, the GFM tolerates CSI NMSE levels 0.5~dB to 4~dB higher than those required by the baselines.} 
In \figref{fig:id_csi}(a), the LMMSE and EP comparison reveals a CSI-error-limited bottleneck, where EP's performance degrades with inaccurate CSI, whereas \figref{fig:id_csi}(b) is predominantly noise-limited. Despite these distinct challenging conditions, the GFM consistently demonstrates {resilience} in mitigating the impact of channel uncertainty and noise contamination.

\subsection{Out-of-Distribution Robustness Study}

\CheckRmv{
  \begin{figure}[t]
    \centering
    \includegraphics[width=3.0in]{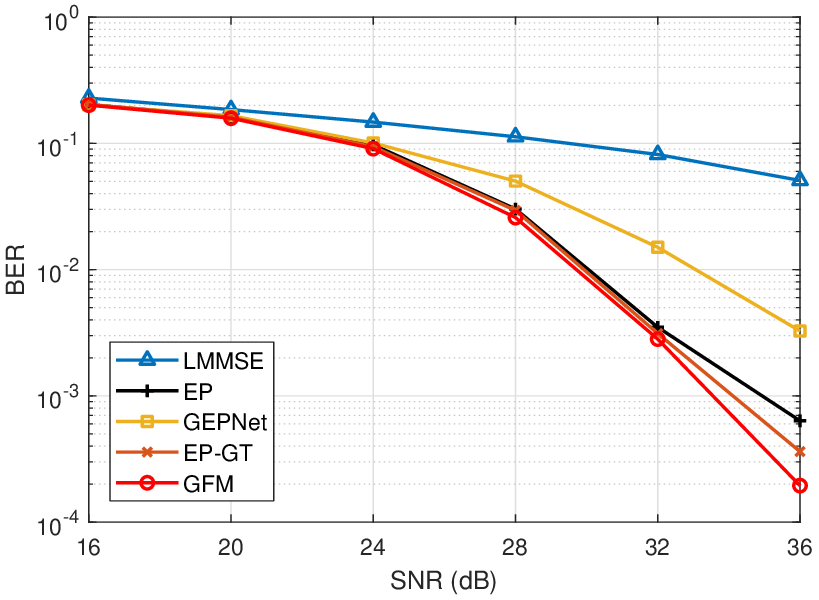}
    \caption{BER performance on the OOD testing dataset D7 featuring the RMa scenario with \Times{96}{32} MIMO and 64-QAM modulation.}
    \label{fig:rma}
  \end{figure}
}

\CheckRmv{
  \begin{figure}[t]
    \centering
    \includegraphics[width=3.0in]{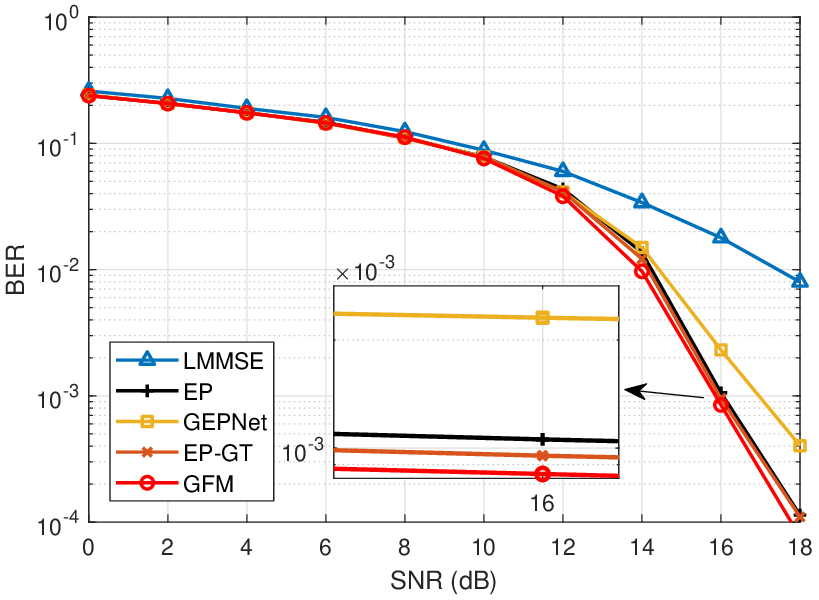}
    \caption{{BER performance on the OOD testing dataset D8 featuring the U6G XL-MIMO RMa system with \Times{1024}{128} MIMO and 64-QAM modulation.}}
    \label{fig:u6g}
  \end{figure}
}

In this subsection, we evaluate the zero-shot performance of the pre-trained GFM on two OOD datasets, D7 and D8, featuring unseen configurations of propagation scenarios, frequency bands, and antenna scales. {We first consider dataset D7, operating under an RMa propagation environment with an antenna ratio of $\nt / \nr = 1 / 3$, which is strictly excluded from the pre-training datasets.}  
As illustrated in \figref{fig:rma}, despite the harsh channel conditioning typical of rural environments, the GFM yields even more pronounced gains over the baselines without any adaptation. This suggests that the model learns transferable interference representations robust to severe distribution shifts. 

{We further investigate the GFM's scalability under the U6G XL-MIMO setup, i.e., dataset D8, which involves a shift to a 6.8 GHz band, an expanded 1024$\times$128 array, and the RMa propagation model. Given its stark deviation from pre-training, this joint domain shift challenges the model's resilience.} As shown in \figref{fig:u6g}, the GFM accommodates the larger array configuration and the variations in high-frequency propagation, maintaining a consistent performance edge over all baselines. These results confirm the proposed GFM's potential to generalize across the tested shifts in propagation scenarios, frequency bands, and array scales. 

\CheckRmv{
  \begin{figure}[t]
    \centering
      \subfigure[{BER versus fine-tuning sample size at \SNR{=}{28}}]{
        \includegraphics[width=3.0in]{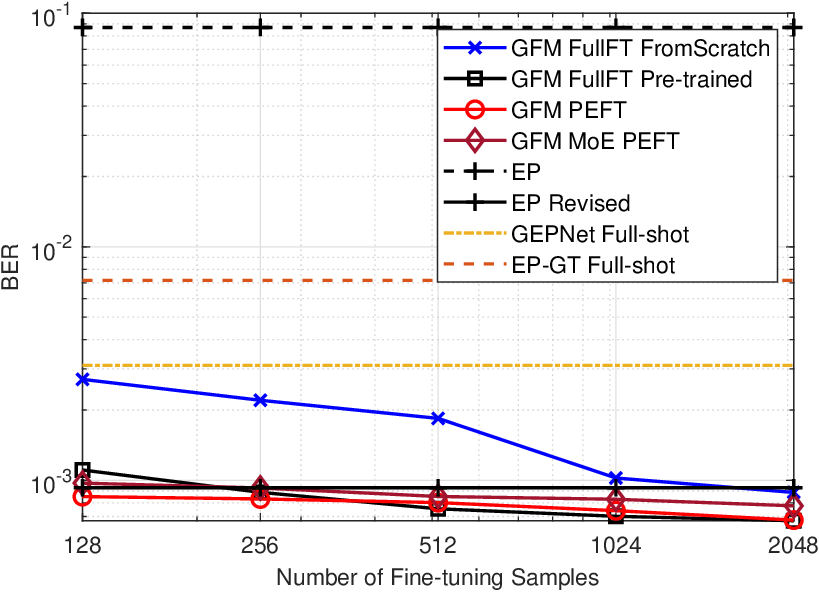} 
        \label{fig:peft_pn_vs_samples}
      }
      \subfigure[{BER versus SNR}]{
        \includegraphics[width=3.0in]{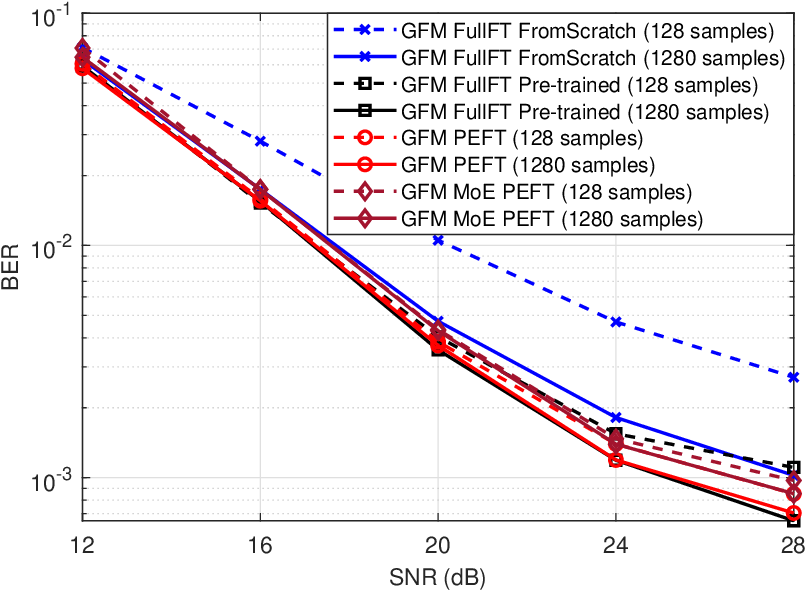}
        \label{fig:peft_pn}
      }
    \caption{{Few-shot adaptation BER performance under the phase noise regime.}}
    \label{fig:peft1}
  \end{figure}
}

\CheckRmv{
  \begin{figure*}[t]
    \setlength{\abovecaptionskip}{-0.1cm}
	  \setlength{\belowcaptionskip}{-0.0cm}
    \centering
    \includegraphics[width=6.2in]{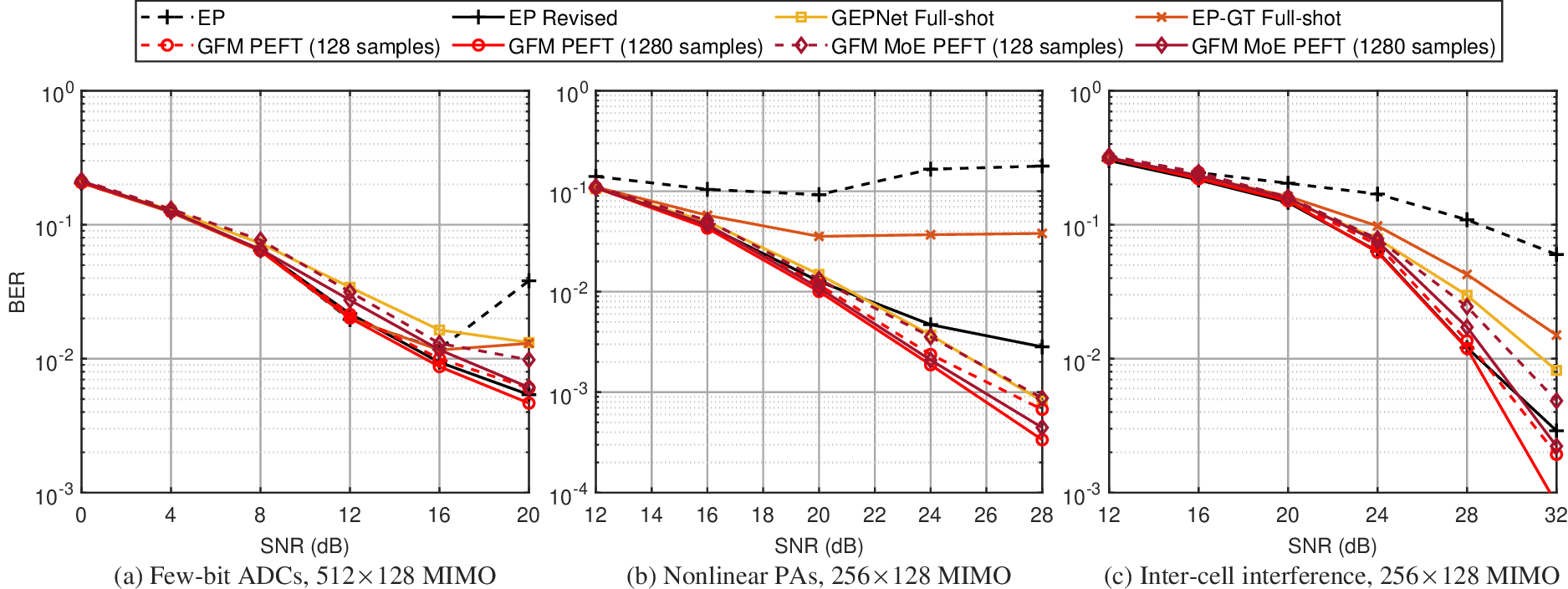}
    \caption{{Few-shot adaptation BER performance across different downstream system regimes with 16-QAM modulation.}}
    \label{fig:peft2}
  \end{figure*}
}

\subsection{Few-Shot Performance across Downstream Regimes}

Beyond evaluating the pre-trained GFM and its zero-shot generalization capabilities, we further investigate the few-shot adaptation performance of the GFM across the representative downstream regimes introduced in Section~\ref{sec:downstream}. 
The comparative analysis 
{involves four distinct adaptation schemes under varying few-shot sample sizes (e.g., 128 and 1280 samples)}
to evaluate the efficacy of the pre-training and fine-tuning strategies: 1) ``GFM FullFT FromScratch'' initializes the model with random weights and optimizes it via full-parameter fine-tuning; 2) ``GFM FullFT Pre-trained'' inherits weights pre-trained on foundation datasets D1--D5 and performs full-parameter fine-tuning; 3) ``GFM PEFT'' also utilizes the pre-trained weights but only fine-tunes the lightweight adaptation modules while freezing the backbone; and 4) ``GFM MoE PEFT'' integrates the condition-aware expert routing mechanism with the PEFT strategy.  
This experimental setup aims to quantify the {value of pre-trained foundation knowledge} in accelerating transfer to distinct regimes under few-shot conditions, and to validate whether the proposed PEFT and MoE can approximate the performance of full-parameter adaptation.
{Moreover, to provide a model-based baseline against mismatches, we introduce an impairment-aware EP (denoted as ``EP Revised''). Assuming knowledge of the impairment parameters, it compensates amplitude non-linearities via Bussgang decomposition, while mitigating multiplicative phase drifts and inter-cell interference through pre-whitening.}

The few-shot adaptation BER performance under the phase noise regime is presented in \figref{fig:peft1}. Specifically, \figref{fig:peft_pn_vs_samples} shows the BER against fine-tuning sample size at a fixed SNR of 28~dB. The figure demonstrates that the proposed PEFT approach converges rapidly with merely 128 fine-tuning samples. This strategy minimizes adaptation overhead by updating only 1.10\% of parameters while approaching full fine-tuning accuracy within approximately 12\%. Furthermore, the performance gap between the ``GFM FullFT FromScratch'' model and the pre-trained GFM {illustrates the benefit} of foundation pre-training, as the from-scratch model struggles to learn meaningful interference representations under such data-scarce conditions. 
{{Notably,} the adapted GFM with PEFT {outperforms} both the statistical ``EP Revised'' and the full-shot learning baselines (GEPNet and EP-GT) trained on all 5,120 fine-tuning samples, proving the superiority of our method in handling phase drifts.}
{The MoE-based scheme also preserves} the rapid convergence of the base GFM while incurring only a slight performance gap relative to the full-activation architecture.
These findings are further corroborated by \figref{fig:peft_pn}, where the ``GFM PEFT'' achieves the most competitive performance across all SNRs using limited fine-tuning samples, while the MoE-based variants exhibit comparable BERs. 

Similar trends are observed for the other regimes in \figref{fig:peft2}. Whether combating few-bit ADCs, nonlinear PA distortions, or colored inter-cell interference, the ``GFM PEFT'' consistently outperforms classical and learning-based baselines using merely 128 training samples (2.5\% of the full dataset),
demonstrating its {adaptability to diverse distortions without requiring prior knowledge of impairment parameters}. {Finally,} the MoE routing dynamically activates experts to reduce inference overhead, as detailed in Section~\ref{sec:complexity}, while sustaining high accuracy given 1280 samples (25\% of the full dataset).

\subsection{Ablation Study}  
\CheckRmv{
  \begin{table*}[t]
    \centering
    \begin{threeparttable}
    \caption{Ablation Study of GFM Components under ID (D6) and OOD (D7) Scenarios (BER Performance)}
    \label{tab:ablation}
    \renewcommand{\arraystretch}{1.2}
    \begin{tabular}{l *{6}{>{\centering\arraybackslash}p{1.2cm}}}
      \toprule
      \multirow{3}{*}{Methods} & \multicolumn{3}{c}{ID: D6, \Times{512}{128} MIMO, 256-QAM} & \multicolumn{3}{c}{OOD: D7, \Times{96}{32} MIMO, 64-QAM} \\
      & \multicolumn{3}{c}{imperfect CSI, \SNR{=}{28}} & \multicolumn{3}{c}{RMa-{NLoS}} \\
      \cmidrule(lr){2-4} \cmidrule(lr){5-7}
      & \multicolumn{3}{c}{NMSE (dB)} & \multicolumn{3}{c}{SNR (dB)} \\
      \cmidrule(lr){2-4} \cmidrule(lr){5-7}
      & $-$29 & $-$25 & $-$21 & 28 & 32 & 36 \\
      \midrule
      GFM (2.19M) & \textbf{2.04e-3} & \textbf{7.69e-3} & \textbf{2.86e-2} & \textbf{2.58e-2} & \textbf{2.82e-3} & \textbf{1.94e-4} \\
      GFM w/o MPNN (2.03M) & 7.03e-3 & 1.44e-2 & 3.25e-2 & 7.51e-2 & 2.76e-2 & 8.58e-3 \\
      GFM w/o GT (0.11M) & 5.76e-3 & 1.46e-2 & 3.16e-2 & 7.00e-2 & 2.78e-2 & 8.65e-3 \\
      \bottomrule
    \end{tabular}
    \begin{tablenotes}
        \footnotesize
        \item Note: (\textasciitilde) denotes the number of parameters, and bold text indicates the best results. The parameter counts do not sum up linearly ($\text{0.11}+\text{2.03}\neq \text{2.19}$) because the input dimension of the decoder heads shrinks when either the GT or MPNN branch is removed (bypassing concatenation).
      \end{tablenotes}
    \end{threeparttable}
  \end{table*}
}

To evaluate the individual contributions of the components within the GFM in pre-training, we conduct an ablation study by comparing the full GFM with two variants: 1) \textit{GFM w/o MPNN}, which removes the MPNN branch, and 2) \textit{GFM w/o GT}, which excludes the GT blocks. 
The BERs of these variants on both the ID dataset D6 with varying CSI uncertainty and the OOD dataset D7 with varying SNRs are summarized in \tabref{tab:ablation}. As observed, the full GFM achieves the lowest BER across all scenarios, while the removal of each component leads to a noticeable performance degradation, {confirming the necessity} of integrating local and global feature extraction.

\CheckRmv{
  \begin{table*}[t]
    \centering
    \begin{threeparttable}
      \caption{Ablation Study of PEFT Components across Downstream Regimes (BER at ${\text{SNR} =28~\mathrm{dB}}$)}
      \label{tab:peft_ablation}
      \renewcommand{\arraystretch}{1.3}
      {\begin{tabular}{l cccccccc}
        \toprule
        \multirow{2}{*}{Scenario} & Full FT & GT LoRA & MPNN Ada. & Dec. Ada. & \shortstack{GT LoRA \\ MPNN Ada.} & \shortstack{GT LoRA \\ Dec. Ada.} & \shortstack{MPNN Ada. \\ Dec. Ada.} & \shortstack{GT LoRA \\ MPNN Ada. \\ Dec. Ada.} \\
        & (100\%) & (0.96\%) & (0.07\%) & (0.07\%) & (1.03\%) & (1.03\%) & (0.14\%) & (1.10\%) \\
        \midrule
        Phase noise & \textbf{6.52e-4} & 7.70e-4 & 4.60e-3 & 9.22e-4 & 8.14e-4 & 7.40e-4 & 8.51e-2 & \underline{7.04e-4} \\
        Nonlinear PAs & \textbf{3.00e-4} & 5.04e-4 & 7.22e-2 & 6.92e-4 & 3.45e-4 & 5.25e-4 & 5.79e-4 & \underline{3.36e-4} \\
        \bottomrule
      \end{tabular}}
      \begin{tablenotes}
        \footnotesize
        \item Note: (\textasciitilde\%) denotes the ratio of trainable parameters within the GFM. ``Dec.'' denotes the decoder heads, while ``Ada.'' stands for the Adapter module. Bold text indicates the top results, whereas the second-best results are underlined.
      \end{tablenotes}
    \end{threeparttable}
  \end{table*}
}

To evaluate the impact of the PEFT strategies for each component within the GFM (\figref{fig:peft}), we investigate the BER performance of various PEFT combinations across representative downstream regimes. 
The ablation results for the phase noise and nonlinear PA regimes at \SNR{=}{28} are summarized in \tabref{tab:peft_ablation}. The comparison reveals that the joint adaptation of all modular components, i.e., GT LoRA, MPNN Adapter, and Decoder Adapter, is essential for achieving the best performance, as the removal of any component leads to a {performance drop}. This ``Full PEFT'' configuration (the last column) outperforms all other PEFT combinations and approaches the performance of full fine-tuning within approximately 12\% while updating only 1.10\% of the total parameters, {indicating the efficiency} of the proposed PEFT design.

\subsection{Computational Complexity Analysis} \label{sec:complexity}  
\CheckRmv{
  \begin{table}[t]
    \centering
    \begin{threeparttable}
    \caption{Average Inference Complexity (FLOPs) per Symbol Vector Detection}
    \label{tab:complexity}
    \begin{tabular}{lcc}
      \toprule
      \multirow{2}{*}{Methods}  & Nonlinear PAs  & Few-bit ADCs \\
       & \Times{256}{128} MIMO &  \Times{512}{128} MIMO\\
      \midrule
      EP & 0.069G  & 0.138G \\
      GEPNet (0.11M) & 9.102G & 9.171G \\
      EP-GT (2.03M) & 2.252G & 2.404G\\
      {GFM {(2.19M)}} & 2.665G & 2.734G \\
      {{GFM w MoE} {(2.19M)}} & {0.450G}  & {0.550G} \\
      \bottomrule
    \end{tabular}
    \begin{tablenotes}
        \footnotesize
        \item Note: (\textasciitilde) denotes the number of parameters.
      \end{tablenotes}
    \end{threeparttable}
  \end{table}
}

In this subsection, we investigate the computational complexity of the proposed detectors compared to the baselines in terms of the floating-point operations (FLOPs) required per symbol vector detection, averaged over 200 channel uses. This metric is recorded using the PyTorch profiler under two representative regimes with distinct MIMO sizes, as shown in \tabref{tab:complexity}.
While the base GFM exhibits lower complexity compared to the GEPNet, activating all model parameters for every instance remains computationally intensive. In contrast, the integration of the dynamic MoE routing mechanism addresses this by conditionally activating only the most relevant expert module (MPNN or GT) based on the perceived task difficulty. 
{As quantified in \tabref{tab:complexity}, the \textit{GFM with MoE} approach yields an 80--83\% reduction in FLOPs relative to its full-activation counterpart. 
Specifically, for the \Times{256}{128} (\Times{512}{128}) system, the standalone MPNN and GT experts require 0.388G (0.457G) and 2.452G (2.517G) FLOPs, respectively; the router adaptively activates the lightweight MPNN expert in 97.0\% (95.5\%) of the instances, with the remaining routed to the GT expert.
Although a gap remains when compared to EP (roughly 4$\times$ to 6.5$\times$ FLOPs), the MoE mechanism effectively bridges the efficiency divide.} 
These results, combined with the performance gains observed in previous subsections, confirm that the proposed framework achieves a desirable trade-off between detection accuracy and computational efficiency.

\section{Conclusion} \label{sec:conclusion}
In this paper, we introduced a wireless-native GFM for large-scale MIMO detection, using a hybrid architecture that unifies localized message passing with global Transformer attention. By iteratively infusing physical interference information derived from the EP algorithm, this physics-informed architecture achieves universal interference representations and precise symbol detection via large-scale pre-training.
{Additionally,} PEFT strategies were designed to resolve the generalization bottleneck through adaptability, enabling rapid transfer to heterogeneous deployment regimes with minimal overhead, while an embedded MoE mechanism dynamically activates only the necessary modules to reduce inference cost. 
{Numerical results demonstrated that} the proposed GFM {consistently} outperforms both classical and data-driven baselines across diverse zero-shot and few-shot conditions.

\appendices



\ifCLASSOPTIONcaptionsoff
  \newpage
\fi


\bibliographystyle{IEEEtran}      
\bibliography{ref_new}

\end{document}